# Graphene MEMS and NEMS

Xuge Fan[1,2,3*], Chang He[1], Jie Ding[3*], Qiang Gao[1], Hongliang Ma[1,3], Max C. Lemme[4,5] and Wendong Zhang[6,7*]

[1]Advanced Research Institute of Multidisciplinary Sciences, Beijing Institute of Technology, 100081 Beijing, China.

[2]Center for Interdisciplinary Science of Optical Quantum and NEMS Integration, Beijing Institute of Technology, 100081 Beijing, China.

[3]School of Integrated Circuits and Electronics, Beijing Institute of Technology, 100081 Beijing, China.

[4]Chair of Electronic Devices, Faculty of Electrical Engineering and Information Technology, RWTH Aachen University, Otto-Blumenthal-Str. 25, 52074 Aachen, Germany.

[5]AMO GmbH, Otto-Blumenthal-Str. 25, 52074 Aachen, Germany.

[6]State Key Laboratory of Dynamic Measurement Technology, North University of China, Taiyuan 030051, China.

[7]National Key Laboratory for Electronic Measurement Technology, School of Instrument and Electronics, North University of China, Taiyuan 030051, China.

*Email: xgfan@bit.edu.cn; jie.ding@bit.edu.cn; wdzhang@nuc.edu.cn

**Abstract**

Graphene is being increasingly used as an interesting transducer membrane in micro- and nanoelectromechanical systems (MEMS and NEMS, respectively) due to its atomical thickness, extremely high carrier mobility, high mechanical strength and piezoresistive electromechanical transductions. NEMS devices based on graphene feature increased sensitivity, reduced size, and new functionalities. In this review, we discuss the merits of graphene as a functional material for MEMS and NEMS, the related properties of graphene, the transduction mechanisms of graphene MEMS and NEMS, typical transfer methods for integrating graphene with MEMS substrates, methods for fabricating suspended graphene, and graphene patterning and electrical contact. Consequently, we provide an overview of devices based on suspended and nonsuspended graphene structures. Finally, we discuss the potential and challenges of applications of graphene in MEMS and NEMS. Owing to its unique features, graphene is a promising material for emerging MEMS, NEMS and sensor applications.

**Keywords:** graphene; MEMS; NEMS; sensors; 2D materials

## 1. Introduction

Graphene was experimentally discovered in 2004 [1,2] and has since attracted substantial attention in fundamental and applied research by the physics, chemistry and materials science communities. Monolayer graphene consists of a one-atom thick sheet of tightly packed carbon atoms that are bonded together in a hexagonal honeycomb lattice. Monolayer graphene is the thinnest known material (approximately 0.335 nm thick) and is extremely strong, chemically stable, and exceptionally conductive [3]. Owing to its high thickness and unique mechanical and electrical properties [4–6], graphene is an extremely promising material for use as an electromechanical transducer in nanoelectromechanical systems (NEMS); as a electromechanical transducer, graphene could substantially reduce the device footprint while providing improved device performance. Furthermore, in monolayer graphene, all the atoms are directly exposed to the environment, thereby impacting the electrical properties of all the atoms of the monolayer graphene; thus, graphene is especially interesting for sensing applications. In addition, because of its planar geometry, graphene is easily compatible with standard lithographic processing. An obvious advantage is that although graphene is intrinsically nanoscale, it can be patterned via standard lithographic processes at the wafer scale.

Graphene is available in the form of large sheets and can be placed on top of various surfaces. Furthermore, graphene can be self-suspended in atomically thick membranes or beams. In addition, graphene can be suspended together with polymers or other membranes. Compared with other carbon-based nanomaterials, such as carbon nanotubes, a prominent advantage of using graphene for practical device applications is that the widely available planar semiconductor processing infrastructure can be used for manufacturing graphene devices. Many practical applications of graphene in areas such as electronics (e.g., transistors, ultrathin and optically transparent electrodes, supercapacitors, and spintronics) [7–9], photonics (e.g., photodetectors and optical modulators) [7,10], and composite materials (e.g., polymer reinforcements, lubricants, and coolants) [7,11–13] have been explored in terms of the unique electrical, optical and mechanical properties of graphene.

Because of its unique properties, graphene is also an extremely attractive material for micro- and nanoelectromechanical systems (MEMS and NEMS, which, in this review, are collectively referred to as "NEMS"). Initially, graphene was used in NEMS for fundamental studies on resonant structures [14] and in gas sensors[15]. Graphene has only recently been widely studied in different NEMS and sensor applications[16,17]. We believe that there are many opportunities to exploit the unique properties of graphene in NEMS devices. Graphene-based NEMS resonators were provided in review papers [18–20,] and brief summaries of graphene in MEMS and NEMS were given in the literature [21,22]. In addition, NEMS sensors based on suspended 2D materials were comprehensively reviewed in 2020[23]. Graphene-based photonic detectors have been comprehensively reviewed recently [24,25] and are not included here.

Compared with previous work, this review paper provides a broad overview of various graphene-based NEMS and devices that have been reported in the literature. We discuss the relevant properties of graphene and present key manufacturing technologies for enabling graphene-based NEMS structures. First, we show different graphene materials, their properties, and different transduction mechanisms for NEMS devices. Consequently, we discuss various graphene manufacturing and integration techniques for NEMS applications, including methods for transferring and patterning graphene, electrical contacting, and realizing suspended graphene. Finally, we summarize different MEMS and NEMS devices based on suspended and nonsuspended graphene structures, such as various types of pressure sensors; resonators; switches; mass sensors; accelerometers; microphones and loudspeakers; gas sensors; hall sensors; and flexible devices, including transparent electrodes, strain sensors, pressure sensors, and audio emitting and receiving devices.

## 2. Graphene properties

Graphene is a two-dimensional material consisting exclusively of carbon atoms that are densely packed in a regular atomic-scale hexagonal pattern. Each atom in a graphene sheet has four bonds: one σ bond with each of its three neighbors and one π-bond that is oriented out of plane. The atoms in graphene are spaced approximately 0.142 nm from each other (i.e., the length of the molecular bonds between the carbon atoms), and the thickness of a layer of graphene is approximately 0.34 nm (i.e., the spacing between two graphene sheets when stacked on top of each other). In general, graphene is a monolayer, bilayer or multiple layers (n<10) of carbon atoms bonded by strong covalent bonds within a layer ($sp^2$ hybridization in a hexagonal lattice and partial filling of π-orbitals above and below the plane of the sheet) with weak van der Waals bonds between neighboring layers. Many reported properties of graphene are exceptional [26], including an intrinsic Young's modulus of 1 TPa [5], an intrinsic strength of 130 GPa [5], a room-temperature electron mobility of up to $2.5 \times 10^5$ $cm^2V^{-1}s^{-1}$ [27], a high thermal conductivity of above 3000 W $mK^{-1}$ [28], an excellent transparency and uniform optical absorption of ≈ 2.3% in a wide wavelength range [29], an impermeability to gases [30], a stretchability of up to approximately 20% [31], and a high compatibility with high densities of electric current [32]. Furthermore, graphene can strongly adhere to $SiO_2$ surfaces [33] and can stably exist in suspended membranes even with a thickness of one atom layer[34]. In addition, graphene can be readily chemically functionalized[35]. Therefore, graphene is a promising material for NEMS applications. Notably, some of these extreme properties can be measured only in high-quality graphene samples, such as mechanically exfoliated graphene without grain boundaries [1] or graphene on hexagonal boron nitride [27,36].

Different methods are used to produce graphene; they include mechanical exfoliation [1,2,14], liquid-phase exfoliation [37–40], chemical vapor deposition (CVD) [41–43] and epitaxial methods [44,45]. The mechanical exfoliation of very high-quality graphene flakes is not easily scalable, and the placement of the resulting μm-scale flakes cannot be adequately controlled for large-scale device applications. For many practical MEMS and sensor applications, CVD graphene is one of the most attractive options. CVD graphene is typically deposited on a support substrate, such as a Cu or Ni foil, from which CVD graphene can be transferred to a variety of other substrate surfaces. CVD graphene on Cu or Ni foils is relatively inexpensive and widely available from commercial suppliers, such as the Beijing Graphene Institute (China), Graphenea Nanomaterials (Spain), Graphene-Supermarket (USA), 6Carbon Technology (Shenzhen) (China), and many others (e.g., 2D Layer (USA), 2D Semiconductors (USA), etc.). A variety of substrate sizes are available (e.g., 10 × 10 mm$^2$ or 100 mm diameter wafer foils), with relatively uniform and high-quality monolayer graphene. CVDs are typically polycrystalline; i.e., they are composed of monocrystalline grains of varying orientations joined by grain boundaries [46–48]. The properties of CVD graphene strongly depend on the material quality; the substrate material on which the graphene sheet is placed; and the graphene grain size, which is typically approximately a few μm to tens of μms[47]. Compared with that of monocrystalline graphene, the electrical properties of polycrystalline graphene may be strongly impacted by the grain boundaries. And electrical and mechanical properties depending on the lattice may not be present in CVD graphene samples that consist of grains and grain boundaries. Typical technologically relevant material properties of polycrystalline CVD graphene are listed in **Table 1** [49], along with the best reported material properties of high-quality graphene [26]. Other available forms of graphene include 1) graphene sheets, which are grown on SiC substrates (i.e., grown via epitaxy); 2) graphene flakes in liquid solution ranging from nm to μm, which are produced through liquid-phase exfoliation and can be coated on the surface, resulting in the coating with dispersed, randomly overlapping graphene flakes; and 3) reduced graphene oxide (GO), which is produced via laser scribing (this process results in loosely stacked graphene layers, graphene aerogels, and graphene foams [26,50]).

**Table 1. Relevant properties of graphene for MEMS and NEMS applications**

| Graphene / Typical Properties | Polycrystalline CVD Graphene | Exfoliated Graphene |
|---|---|---|
| Young's Modulus | 1.02 TPa [51] | 1.02 TPa [5] |
| Fracture Strain | Up to 20% [31] | Up to 30% [5,52,53] |
| Electrical Conductivity | On SiO$_2$: 4050 cm$^2$/Vs [42]<br>On h-BN: 350000 cm$^2$/Vs [54]<br>Suspended: 15000 cm$^2$/Vs [55] | On SiO$_2$: 15000 cm$^2$/Vs [2]<br>On h-BN: 275000 cm$^2$/Vs; (T= RT): 120000 cm$^2$/Vs [56]<br>Suspended: 185000 cm$^2$/Vs [57] |
| Sheet Resistance | 72 Ω/sq [58]<br>Few hundred Ω/sq to > 10$^4$ Ω/sq [59] | Few hundred Ω/sq to > 10$^4$ Ω/sq [59] |
| Thermal Conductivity | Suspended: ~1500-5000 W/mK [60] | Suspended: ~4840-5300 W/mK [61,62] |
| Temperature Coefficient of Expansion | 7 × 10$^{-6}$ K$^{-1}$ [63] | 7 × 10$^{-6}$ K$^{-1}$ [64] |
| Permeability | Gases or ions might permeate grain boundaries or defects; therefore, multilayer CVD graphene is better [65] | Impermeable, even against helium [30] |
| Gauge Factor | 3 (suspended drum) [34]<br>>300 (nanographene films) [66] | 1.9 (suspended beam) [67]<br>150 (on SiO$_2$) [68] |

## 3. Transduction mechanisms of graphene for NEMS

Many transduction mechanisms in graphene can be used in NEMS and sensor devices; these mechanisms include the piezoresistive effect in graphene, electrostatic sensing (including capacitive and transconductance change by a gate electrode), electrostatic actuation with graphene electrodes, resonant and optical transduction, Hall effects, and resistance changes in graphene due to surface interactions. Notably, electrical signals from graphene is extremely sensitive to various environmental parameters. Hence, parameters such as small changes in the humidity of the surrounding air in contact with the graphene [69–72], gases[70,71,73], temperature[74] or electromagnetic fields from nearby circuits impact the electronic signals of graphene devices. Therefore, to ensure the reliability of electrical signals from practical graphene-based device applications, all critical environmental parameters must be controlled, and the graphene must be carefully shielded from environmental influences, such as electromagnetic fields and changes in environmental gas composition. For example, graphene can be encapsulated by h-BN, $Si_3N_4$, $Al_2O_3$, or poly(methyl methacrylate) (PMMA)-based polymers. Each of the transduction mechanisms is discussed in the following subsections.

### 3.1 Piezoresistive transduction

Graphene has been demonstrated to have a piezoresistive effect; i.e., the mechanical strain induced in graphene results in a change in the graphene resistance[75,76]. Two types of basic functional structures based on the piezoresistive effect of graphene are shown in **Fig. 1**a, b[34,66]. The first type is a suspended graphene membrane that can act at the same time as the membrane and strain gauge (**Fig. 1**a)[34]. The second type is a graphene strain gauge that can be placed on a substrate or a membrane (**Fig. 1**b)[66,77]. For exfoliated and CVD graphene with typical grain sizes of a few μm, piezoresistive gauge factors, which are the ratios of the relative resistance change to the mechanical strain, have commonly been reported to be approximately 1.9 to 8.8 (**Fig. 1**c-f)[67,78–82]. In such graphene sheets, the resistance change of graphene is caused by external strain–induced changes in the electronic band structure in graphene. Theoretical analysis and reported measurements suggest that the direction of the strain in relation to the graphene crystal lattice orientation has no significant influence on the gauge factor, and uniaxial and biaxial strain in graphene result in similar gauge factors (**Fig. 1**g)[83]. Furthermore, the reported piezoresistive gauge factors of bilayer graphene structures are similar to those of monolayer graphene (**Fig. 1**h)[84]. However, the piezoresistive properties of graphene are associated with the graphene type, the graphene quality (e.g., gain size and density), the structure (e.g., suspended graphene membrane and graphene placed on a substrate), and the materials that are in contact with the graphene (e.g., substrate and passivation layers).

There have been isolated reports of very high gauge factors in graphene; these factors range from 150 [68] to 17980 [85]. However, the exact graphene materials used were not specified in the original or follow-up studies, which yielded comparable results. Here, the piezoresistive effect of graphene is caused by mechanical strain in the graphene, which changes the electronic band structure. The piezoresistive behavior of graphene is a superposition of carrier density and carrier mobility modification, where the latter dominates. The piezoresistive behavior is independent of the crystallographic orientation, and the gauge factor is independent of the doping concentration. For small strains, the intrinsic piezoresistive gauge factor of graphene is largely invariant in strain magnitude (**Fig. 1**i) [34,78,86,87].

Large gauge factors have also been reported for coatings with randomly overlapping graphene flakes, for loosely stacked graphene layers prepared by reducing GO via laser scribing, and for graphene-based composites and graphene films on soft substrates. For example, gauge factors of over 300 can be obtained from synthesized nanographene films (which comprise densely packed nanographene islands) by controlling the piezoresistive response of these films by changing their growth parameters [66]. Additionally, gauge factors of more than 600 can be achieved from nanographene films [88]. Graphene-based thin-film strain gauges that produce solution-processed overlapping graphene flakes via spray deposition have high and tunable gauge factors with maximum values greater than 150 [89]. A high gauge factor of ~ 112 for porous graphitic structures can be achieved by direct laser writing from polyimide **(Fig. 1j)**[90]. A gauge factor of approximately 11.4 for a graphene/epoxy composite within the range of 1000 microstrains has been reported [91]. More interestingly, depending on the strain level, high gauge factors ranging from $10^3$ to $10^6$ for graphene woven fabrics, including woven graphene microribbons, have been reported because of their woven mesh configuration, high-density crack formation, and propagation mechanical deformation [92]. A strain sensor made of a vinyl-ester polymer composite filled with multilayer graphene nanoplatelets has a gauge factor of ~ 45 [93], whereas percolative film devices based on N-methyl-2-pyrrolidone (NMP)/graphene flakes on flexible plastic substrates have gauge factors ranging from 10 to 100 [89].

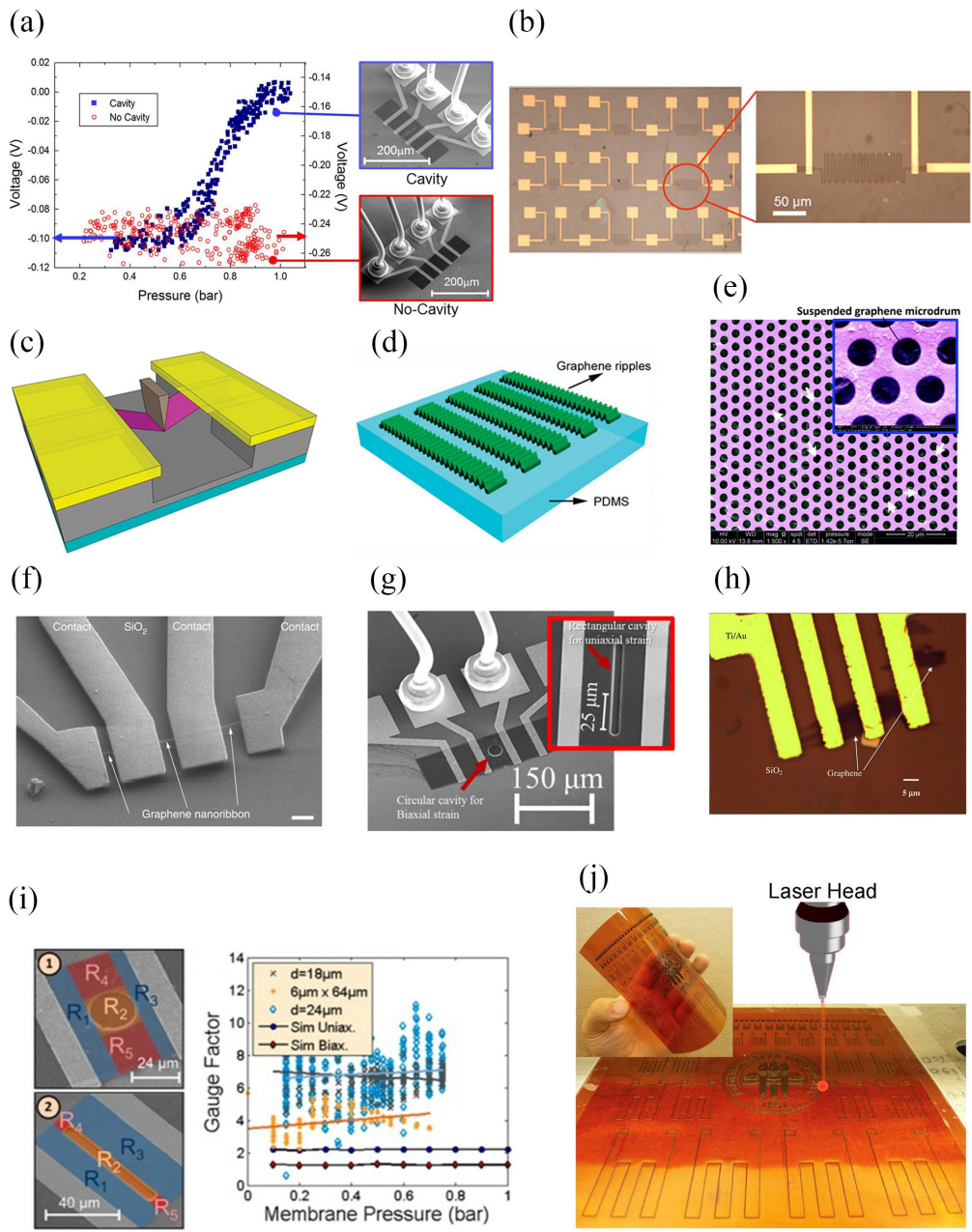

**Fig. 1 Piezoresistive transduction of graphene. a** A piezoresistive NEMS pressure sensor based on a suspended monolayer graphene membrane[34]. **b** Nanographene-based strain sensor devices[66]. **c** Nanoindentation experiments performed on suspended graphene devices for measuring strain and gauge factors[67]. **d** Buckled graphene ribbons on a Polydimethylsiloxane (PDMS) substrate[80]. **e** Arrays of suspended graphene microdrums[81]. **f** A graphene nanoribbon device[82]. **g** Biaxial strain in suspended graphene membranes[83]. **h** An optical microscope image

of a monolayer graphene sheet on a SiO$_2$ surface for measuring gauge factors[84]. **i** Extracted gauge factors of both uniaxially and biaxially strained graphene membranes[87]. **j** Direct laser writing–fabricated graphitic sensor patterns on a polyimide film[90]. Figures reproduced with permission from Refs. 34, 66, 67, 80-84, 87, and 90.

**3.2 Electrostatic transduction**

Electrostatic sensing commonly includes both capacitance changes between moving and fixed plates as their distance and position are changed or media are replaced and transconductance changes by a gate electrode. A high capacitance sensitivity of the graphene membrane for pressure sensing was achieved [94]; this sensitivity is 770% greater than that of frequently used silicon-based membranes. Graphene field-effect transistors are typical devices based on transconductance changes caused by gate electrodes. Graphene, with its novel and electron–hole symmetric band structure and high carrier mobilities and thermal velocities, has gained much attention for overcoming the limitations of Moore's law and 'beyond CMOS' devices. Although pristine monolayer graphene is a gapless semiconductor and current control by gate electrodes is challenging, off-state leakage currents are high, and the current does not readily saturate with the drain voltage. Several reviews on graphene-based field-effect transistors have been published in recent years [95–98].

One of the most popular actuation mechanisms used in MEMS/NEMS is electrostatic actuation, which is realized by the electrostatic force (attraction) between moving and fixed plates by applying a voltage between them. For example, flexible strain sensors made of graphene flakes with electrostatic actuation have a mechanical resonance frequency of approximately 136 MHz under ambient conditions [99]. One typical application of MEMS/NEMS electrostatic actuators is nanoelectromechanical switches, which use electrostatic forces to mechanically deflect the active element into physical contact with an electrode. Graphene has a high Young's modulus, high atomic thickness and high electron mobility, thus making graphene a promising candidate material for NEMS switches. As shown in **Fig. 2**a, the graphene NEMS switch resulted in minimal electrical leakage, a sharp switching response, a low actuation voltage, and a high on/off ratio [100–104]. Typically, the heavily doped silicon substrate is used as the actuation electrode, thus causing the graphene NEMS switch to suffer from a relatively large pull-in voltage (> 10 V) mainly because of the thick, inevitable dielectric gap [105]. To reduce the pull-in voltage, a local actuation gate electrode is normally needed. Both the structures of the graphene cantilever and doubly clamped graphene can be used for graphene MEMS/NEMS switches (**Fig. 2**b-d) [106–110].

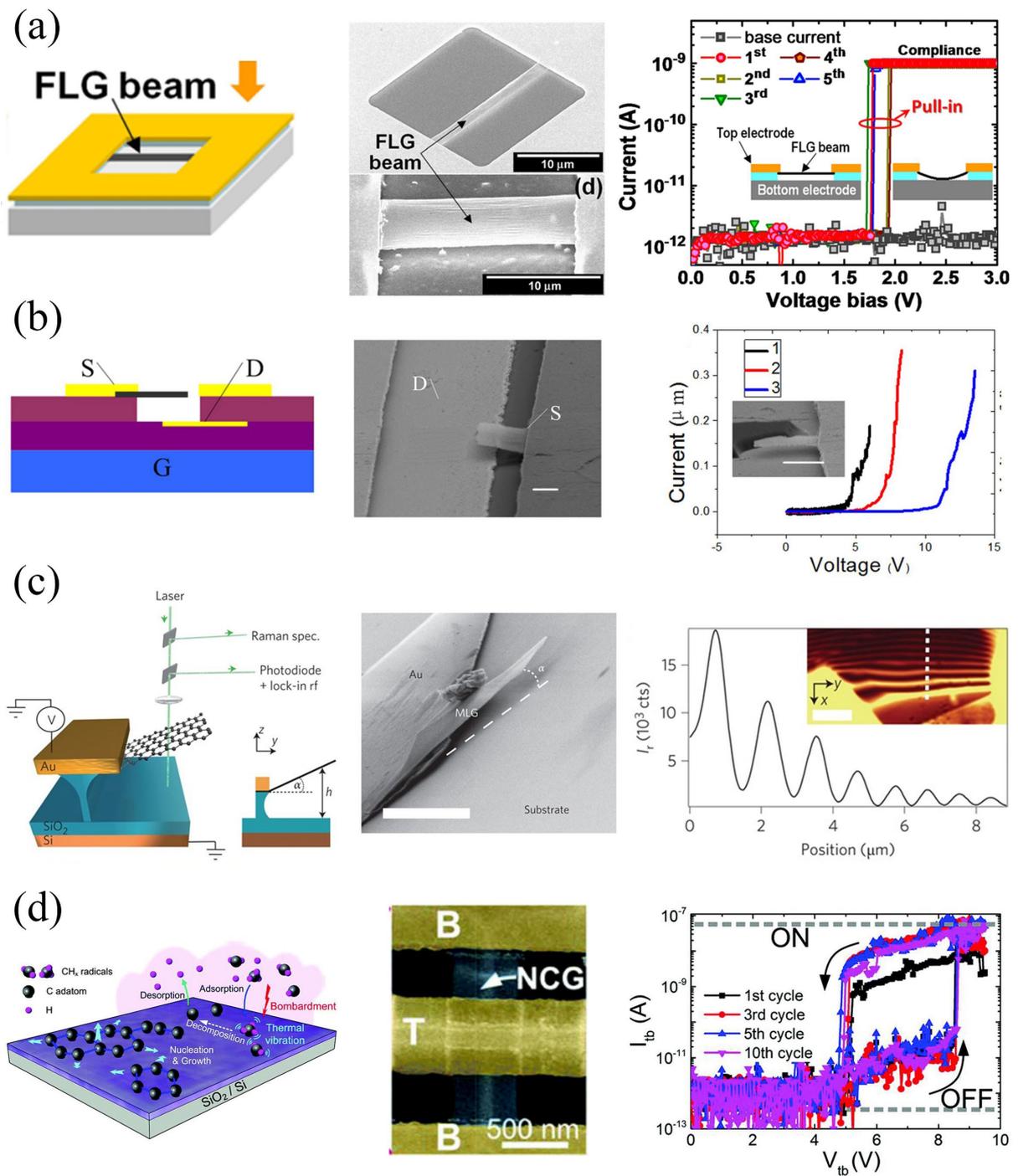

**Fig. 2 Electrostatic transduction of graphene. a** A suspended few-layer graphene beam electromechanical switch[101]. **b** An MLG cantilever overhanging silicon oxide[107]. **c** Nanoelectromechanical switches based on nanocrystalline graphene directly deposited on insulating substrates[110]. **d** A 3-terminal nanoswitch based on a graphene cantilever[106]. Figures reproduced with permission from Refs. 101, 106, 107 and 110.

**3.3 Resonant transduction**

The low areal mass density and high tension of graphene membranes enable graphene-based resonant structures to achieve very high resonance frequencies. Since loudspeakers and microphones can provide only a flat-band response below the lowest resonance mode, graphene speakers[111] and microphones[112] could benefit the transduction of ultrasound. The thermal mechanical pressure noise $p_{noise}$ in a microphone is given by $p_{noise} = \sqrt{\frac{4\,k_B T \sqrt{km}}{A^2 Q}}$, where $k$ denotes the spring constant, $m$ denotes the mass and $A$ denotes the area of the membrane. This equation shows that the increase in noise due to a reduction in the area of the microphone membrane can be compensated for by the reduced mass and spring constant of the graphene membranes.

A method for resonant pressure sensing is based on suspended graphene membranes, which do not require a hermetically sealed cavity; instead, pressure is measured by high-frequency compression of the gas in a 'squeeze film' between the graphene membrane and the substrate[113]. Because of the high compression frequency, the gas has no time to escape even in a nonhermetic cavity with a venting channel[113]. As a result, the trapped gas acts as a pressure-dependent spring that increases the resonant frequency of the graphene membrane.

Optical transduction of graphene NEMS resonators has been used [114–117]. Typically, an intensity-modulated laser is used to periodically heat the suspended graphene via optical absorption, which photothermally drives the motion of the suspended graphene via periodic contraction/expansion. Another laser is used for detection because of the interference caused by the multireflection of light within the graphene resonators[18,19]. Specifically, a laser reflects from the suspended graphene and simultaneously reflects off the substrate after it passes through the suspended graphene. As the suspended graphene vibrates, the space between the suspended graphene and the substrate changes correspondingly, thereby ultimately changing the interferometry condition of the graphene resonators. The displacement and related parameters of the graphene resonator can be obtained by measuring the intensity modulation of the reflected light. The light signal is then converted into an electrical signal by using a fast photodetector. This technique has minimal requirements for the device structure (e.g., not requiring electrodes) and material properties (e.g., no need to be conductive) and is suitable for quickly measuring NEMS resonators made from graphene and emerging 2D materials. Furthermore, this technique is very sensitive to the motion of graphene resonators and can measure the undriven thermomechanical resonance caused by the Brownian motion of the graphene device[14,19].

**3.4 Transduction based on the Hall effect**

The Hall effect refers to the generation of a voltage difference (Hall voltage) across an electrical conductor that is transverse to an electric current in the conductor and a magnetic field perpendicular to the current [118]. The Hall effect is a transducer mechanism whose output voltage varies in response to a magnetic field and can be used to measure the charge carrier density, charge carrier mobility or magnetic field. The Hall element is a simple device structure that can be easily fabricated, and the performance of a Hall element depends mainly on the charge carrier mobility and density of the material [119]. Graphene, as an atomically thin 2D material, has very high charge carrier mobility and low charge carrier density at room temperature and is therefore a promising material for high-performance Hall sensors. As shown in **Fig. 3**, a typical cross-shaped graphene Hall element is a four-terminal device with two pairs of electrodes connected with four terminals of the graphene Hall cross, which are used for supplying current (current mode) or voltage (voltage mode) and measuring the Hall voltage [120]. A graphene Hall sensor can be used in either current mode or voltage mode. According to the Hall effect, when an external magnetic field B is applied perpendicular to the graphene surface and a constant current (I) or a constant voltage (V) is supplied through the graphene channel, a Hall voltage ($V_H$) that is linearly related to the strength of the magnetic field is produced as follows [120–122]:

$$V_H = S_I \times I \times B \qquad (1)$$

$$V_H = S_V \times V \times B \qquad (2)$$

where $S_I$ and $S_V$ denote the current-related and voltage-related sensitivities, respectively, which are the key performance parameters of the graphene Hall elements. A simple derivation showing that the current- and voltage-related sensitivities of a graphene Hall element correlate with carrier density or mobility can be expressed as [120–122]

$$S_I = \frac{1}{qn_s} \qquad (3)$$

$$S_V = \frac{\mu W}{L} \qquad (4)$$

where q denotes the elemental charge; W and L denote the width and length of the graphene channel, respectively; and $n_s$ and $\mu$ denote the sheet carrier density and carrier mobility of the graphene channel, respectively. As Equations (3) and (4) show, the sensitivity of the graphene Hall element can be improved by decreasing the charge carrier density, increasing the charge carrier mobility and optimizing the dimensions of the graphene active region. By changing the carrier density in the current mode, we can easily adjust the sensitivity of the graphene Hall sensors by using a gate voltage and by modifying the surface or substrate [121]. Graphene Hall elements generally feature high sensitivity (**Fig. 3**a, b) [120,123,124]. The combination of high sensitivity and low noise of graphene devices results in the higher resolution of magnetic field (**Fig. 3**c)[125,126]. In addition to silicon-based substrates, graphene Hall elements can also be fabricated on flexible substrates (**Fig. 3**d)[127].

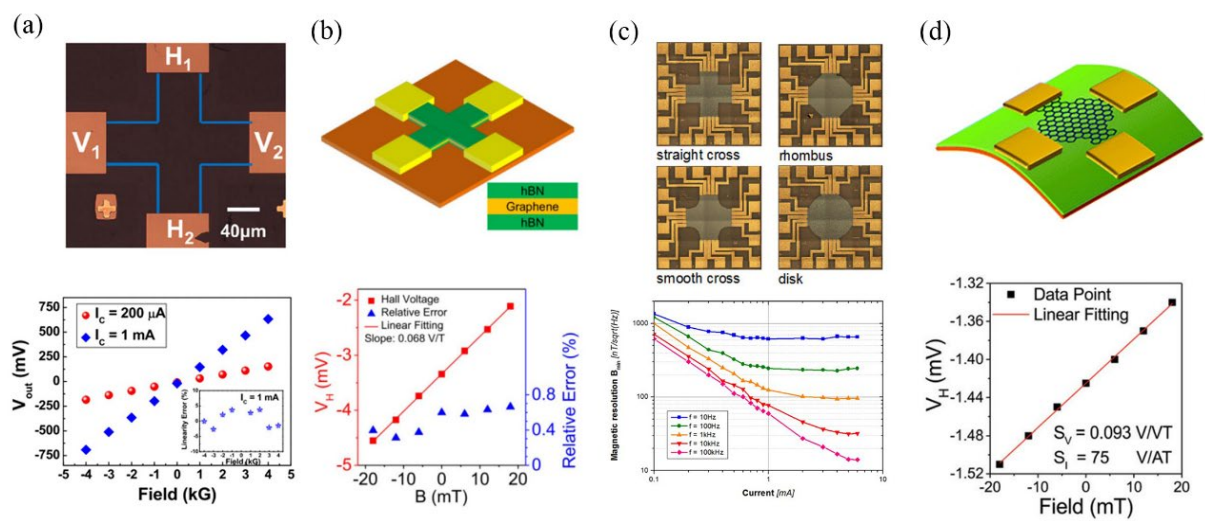

**Fig. 3 Transduction based on the Hall effect. a** A graphene Hall effect sensor[123]. **b** A graphene Hall effect sensor encapsulated by h-BN [124]. **c** Quasifree-standing hydrogen-intercalated monolayer graphene Hall effect sensors with different shapes and room-temperature magnetic resolutions [126]. **d** A flexible Hall sensor based on graphene[127]. Figures reproduced with permission from Refs. 123, 124, 126 and 127.

### 3.5 Other transduction mechanisms

Although unstrained pristine graphene does not exhibit piezoelectric properties, recent reports suggest that graphene exhibits piezoelectric behavior in certain configurations, such as applying strain or engineering graphene; this behavior can be caused by the charge transfer along a work function gradient introduced by the biaxial strain-engineered band structure of the suspended graphene membranes over the cavity [128] or associated with the chemical interaction of graphene's carbon atoms with the oxygen from the underlying $SiO_2$ grating substrates [129]. The piezoelectric transduction of strained or engineered graphene has potential applications in nanogenerators or actuators; however, the extent to which these properties can be used in practical device applications remains to be determined. In addition, suspended graphene can be manipulated by electrostatic and thermal control [130].

Resistance transduction is another transduction mechanism of graphene. Graphene has an exceedingly large surface-to-volume ratio, which enables electronic transport in graphene to be ultrasensitive to the surrounding chemical environment. Adsorption of molecules onto graphene results in the modulation of charge carriers by molecular or vicinity doping, which can be used for humidity and gas sensors [131–133]. The charge transfer by molecular adsorbates impacts the density of states of graphene at the Fermi level; therefore, graphene is capacitated to alter its conductivity [134,135]. Adsorbates behave as donors or acceptors depending on the doping abilities of the introduced molecules, thus resulting in a conductance change in graphene. Appropriate modification or functionalization results in drastic improvement in the sensitivity of graphene-based chemical sensors, and it can also provide guidance for better selectivity or specificity of graphene-based sensors[136].

Recently, a transduction mechanism based on the graphene-induced nonradiative transition for detecting MEMS force and acceleration was studied via analytical solutions and finite element analysis methods[137]. Specifically, the graphene-induced nonradiative transition is highly sensitive to distance, and the deflection of the graphene ribbon is highly susceptible to applied force or acceleration. In addition, a transduction mechanism based on the contact resistance of an improved graphene aerogel for detecting acceleration was reported in 2023[138].

## 4. Transfer methods for integrating graphene with target substrates

Graphene films can be derived from exfoliation, CVD synthesis or epitaxial growth. The transfer of graphene works as a bridge connecting the production of graphene and its applications. To realize the applications of graphene in MEMS/NEMS, graphene typically needs to be transferred from the initial substrate to the target substrate. Concerning the graphene being transferred, having a clean surface and few defects on a large scale is crucial to the performance of graphene devices and is highly important for potential industrial applications of graphene films.

### 4.1 Transferring graphene prepared by mechanical exfoliation

Scotch tape[1] and the tip of an AFM silicon cantilever[139] are two earlier methods for mechanically exfoliating graphene flakes from a thin piece of graphite and simultaneously transferring the graphene flakes to the target substrate. Although mechanical exfoliation produces high quality graphene flakes [140,141], their lateral size is normally limited to a few micrometres, and their thickness is not easily controlled.

**4.2 Transferring graphene prepared via epitaxial growth**

One of the methods used to produce large-scale and high-quality graphene sheets with a single orientation is epitaxial growth on SiC wafers by self-limiting sublimation of Si. Because the commonly used chemical etchants cannot dissolve SiC, traditional wet transfer methods cannot be used for transferring graphene grown on SiC substrates. Graphene grown on a SiC substrate can be transferred onto a target substrate via exfoliation or peeling techniques according to the differential adhesion forces or binding energies between the graphene and different materials. For example, graphene can be transferred from a SiC substrate to a target substrate via either a bilayer film of gold (or palladium) or polyimide (PI), where the deposited metal layer provides the adhesion force and the PI works as a supporting layer during delamination [142,143], or a thermal release tape [144] or Scotch tape [145]. These methods usually cause issues, such as residues, defects or unwanted doping of graphene. A water-soluble polymer poly(vinyl alcohol) (PVA) can be used to transfer graphene from a SiC substrate to a $SiO_2$/Si substrate with PVA dissolved in water; this method avoids the use of a chemical solvent, and thereby, doping-free graphene can be obtained [146]. In short, transferring graphene from a SiC to a target substrate is possible, but achieving high-quality transferred graphene is difficult, thus possibly hindering the widespread use of graphene in NEMSs. Therefore, methods for transferring graphene onto SiC should be further developed.

**4.3 Transferring graphene prepared via CVD**

CVD is an efficient method for synthesizing graphene on a large scale, and Cu is the most commonly used metal substrate for graphene grown by CVD because of the low C solubility of Cu, controllable number of graphene layers and ease transfer at a relatively low cost. Therefore, CVD graphene on a Cu substrate is used as an example to introduce transfer methods for graphene grown by CVD. A face-to-face transfer method for producing wafer-scale graphene films was developed by using nascent gas bubbles and capillary bridges between the graphene film and the underlying substrate during etching of the metal catalyst, thereby accomplishing both growth and transfer steps on one wafer (**Fig. 4**a) [147]. CVD graphene can be transferred via either wet or dry transfer, which are two basic methods.

**4.3.1 Traditional wet polymer-supported transfer of CVD graphene**

Among the wet transfer methods, which are an important way to transfer CVD graphene, polymer-supported transfer methods are the most used to transfer CVD graphene from the metal growth substrate to a target substrate. In most cases, PMMA is used as the polymer supporting layer for wet methods for transferring CVD-grown graphene because of PMMA's relatively low viscosity, excellent wetting capability, flexibility and good dissolubility in several organic solvents [148]. The typical transfer steps are basically as follows: first, the graphene on the back side of the growth substrate (e.g., Cu) is removed by $O_2$ plasma etching; second, the PMMA layer is spin-coated on the graphene and cured on a hotplate; third, the underlying Cu substrate is etched to free the PMMA/graphene layer; fourth, the PMMA/graphene is transferred to the target substrate; finally, the PMMA layer is dissolved in acetone and dried.

The etchants that are commonly used to remove the Cu substrate include aqueous solutions of iron nitrate, iron chloride and ammonium persulfate. After the Cu substrate is removed, the metal particulates that can act as scattering centers to degrade the carrier transport properties are minimally removed by washing and normally require special cleaning technology [149]. However, cracks might easily form during the above graphene transfer because the surface of the Cu substrate is reconstructed at high temperatures during CVD growth and thereby becomes rough. The graphene is grown by following the rough surface of the underlying Cu substrate, and the graphene surface is also rough when the Cu is removed by the etchant. Because the rough surface of the graphene cannot be in full contact with the target substrate, gaps exist between them; consequently, cracks ultimately form during the dissolution of the PMMA layer [150]. To improve the contact between the graphene and the transfer target substrate and reduce the formation of cracks, a second layer of PMMA can be spin coated after the PMMA/graphene stack is transferred to the target substrate [150]. In addition, some water probably remains in the gaps between the PMMA/graphene stack and the target substrate, thus leading to cracks after PMMA is removed. Therefore, it is helpful to improve the quality of the transferred graphene by baking the graphene samples (e.g., at 150 °C for 15 min) to evaporate the water between the graphene and the substrate before removing the PMMA (**Fig. 4**b)[149,151].

Based on the optimized wet PMMA-supported transfer method, a double-layer stacked graphene is created by transferring a monolayer CVD graphene to another monolayer CVD graphene on copper (**Fig. 4**c)[71,152]. The experimental results for this method confirmed that, for suspended structures made of monolayer CVD graphene, the addition of a second layer of CVD graphene disproportionally increases the fracture toughness of the resulting suspended graphene structure and indicated that suspended structures made of double-layer stacked CVD graphene have better overall mechanical resilience than do suspended structures made of monolayer CVD graphene [153–155].

PMMA is not typically removed entirely even after long exposures to acetone because of its high molecular weight and high viscosity, and the PMMA residues might cause charged impurity scattering and unintentional graphene doping [156], thereby affecting the graphene electrical properties, such as increased sheet resistance and related applications (such as in conductive electrodes) [150]. A thermal annealing step in a vacuum furnace under a $H_2$-Ar or $H_2$-$N_2$ environment is often used to remove as many PMMA residues as possible [157]. Notably, methods of annealing are unsuitable for flexible substrates because of the high temperature. There are also other improved methods for removing PMMA residues; these methods include inserting an organic small-molecule buffer layer with excellent solubility in the solvent between the graphene and the PMMA [158] or using UV irradiation to degrade the PMMA [159]. In addition to PMMA, poly(bisphenol A carbonate) (PC) [160], the polymer polyisobutylene (PIB) [161], pentacene ($C_{22}H_{14}$) (**Fig. 5a**)[162], anthracene[163], paraffin[164], photoresist[165], and rosin and/or anisole[166] are also used as supporting layers in wet transfer methods; these supporting layers might be helpful for obtaining clean, less doped and high-quality graphene.

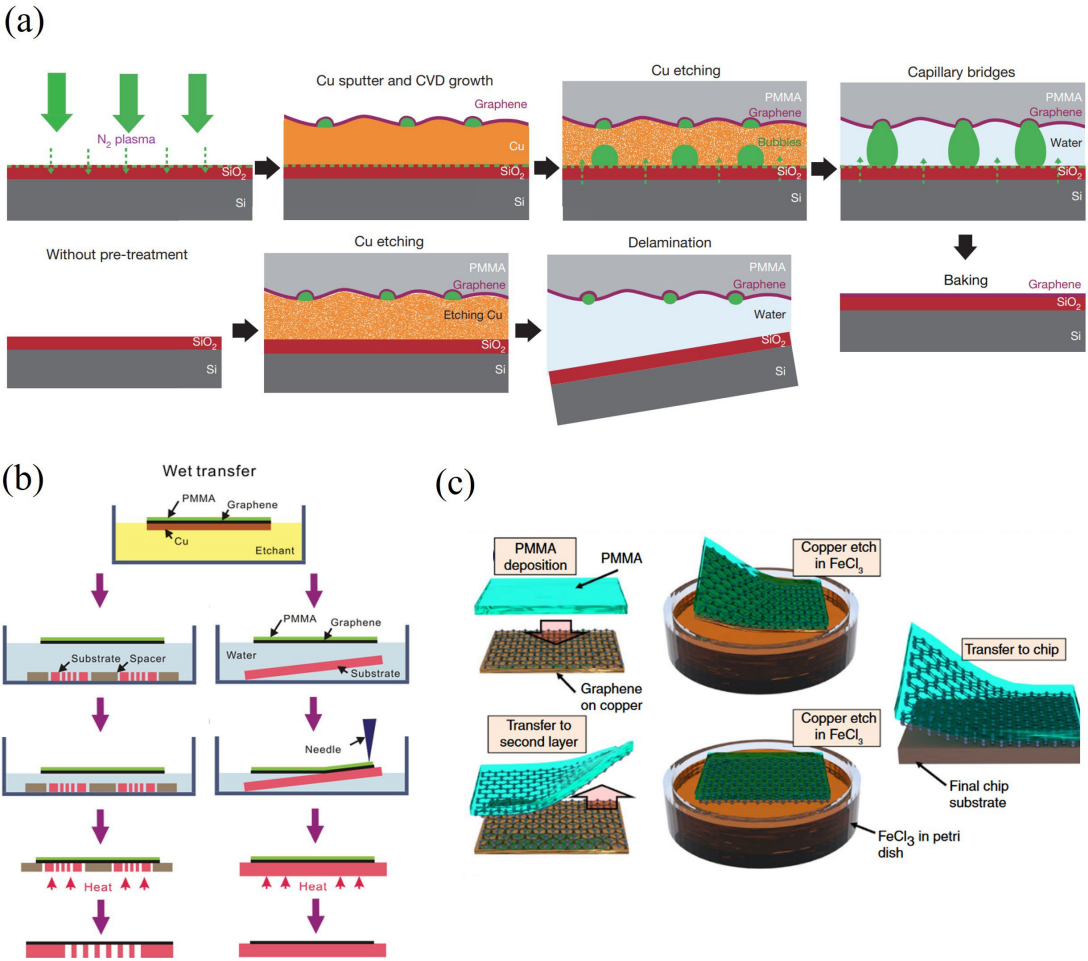

**Fig. 4 Schematics of traditional wet polymer-supported transfer of CVD graphene. a** A face-to-face method for transferring graphene via capillary bridges; the steps shown for this method are as follows: bubbling seeding via plasma treatment, CVD growth, Cu film etching, the formation of capillary bridges and the removal of water and PMMA[147]. **b** and **c** Different methods of PMMA-based wet transfer of CVD-grown monolayer graphene[151,154]. Figures reproduced with permission from Refs. 147, 151 and 154.

### 4.3.2 Thermal release tape for wet transfer of CVD graphene

Many extended methods for transferring CVD graphene are based on basic polymer-supported transfer methods. For example, a thermal release tape can be used as an alternative to the polymers mentioned above and is often used to transfer graphene onto flexible substrates[167]. As shown in **Fig. 5**b, the thermal release tape method has been used to demonstrate roll-to-roll production of 30-inch graphene films for transparent electrodes[168]. A universal method for preparing large-area suspended mono- and few-layer graphene and other 2D materials was developed by using a thermal release tape to transfer graphene onto a densely patterned hole array substrate that is pretreated either with oxygen plasma or via gold film deposition (**Fig. 5**c)[169]. A method for achieving a clean and less defective transfer of monolayer graphene from copper foil onto the target substrate by floating the copper foil with the graphene in hot water and delaminating the graphene by using a thermal release tape has been reported[170].

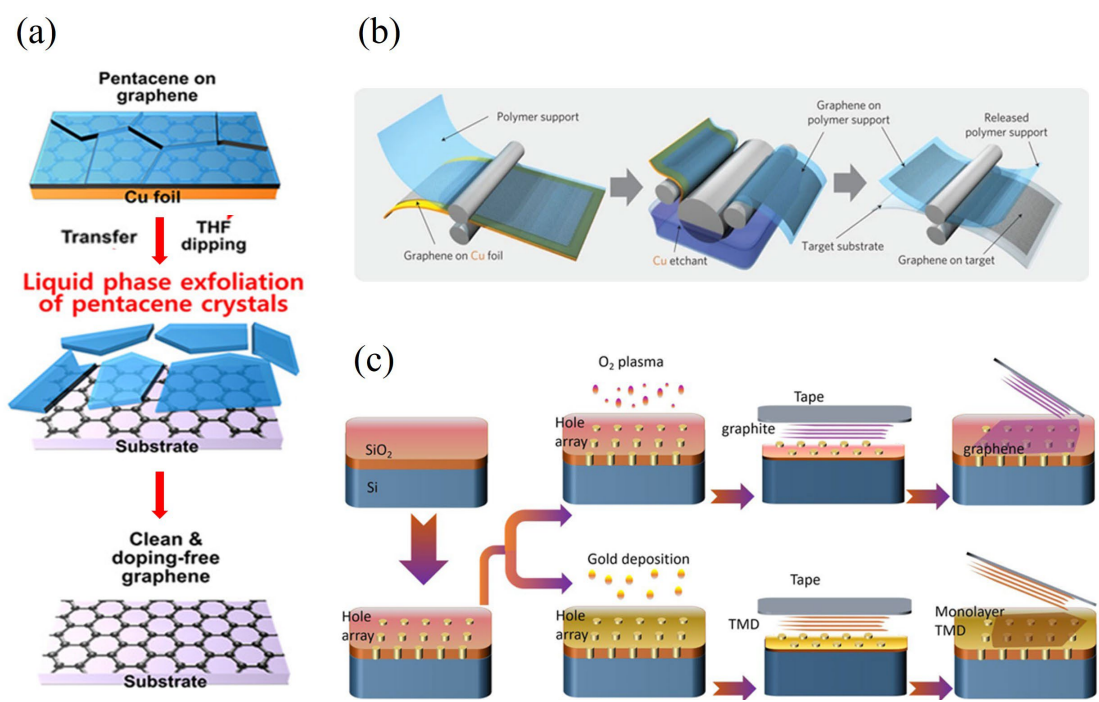

**Fig. 5 Schematics of several typical methods for transferring CVD graphene. a** A clean transfer procedure for transferring wafer-scale graphene from copper foil onto an arbitrary substrate by using pentacene as the supporting layer[162]. **b** A roll-to roll transfer of 30-inch graphene films onto a copper foil[168]. **c** An efficient mechanical exfoliation method for transferring graphene and other 2D materials onto etched cavities[169]. Figures reproduced with permission from Refs. 162, 168 and 169.

**4.3.3 Electrochemical delamination for the wet transfer of CVD graphene**

Electrochemical delamination (also called "bubble transfer") is a fast transfer method for CVD graphene grown on a Cu substrate [171,172]. As shown in **Fig. 6**a, the fast delamination of graphene from the Cu substrate is induced by the production of hydrogen bubbles during the electrolysis of water. Electrochemical delamination can transfer graphene on both sides of metal substrates with high efficiency, and the metal substrates can be reused. In addition, electrochemical delamination can also be used to transfer graphene grown on chemical insert or noble metal substrates, such as Ir, Pt and Au; traditional wet graphene transfer methods cannot be used because of their low solubility and high cost. However, hydrogen bubbles produced during electrochemical delamination may lead to mechanical damage to the transferred graphene. Therefore, several other types of nondestructive electrochemical delamination methods without the production of bubbles have been developed [173,174].

**4.3.4 Polymer-free methods for the wet transfer of CVD graphene**

To avoid the polymer residues that are produced by polymer-supported transfer methods and improve the electrical properties (such as mobility) of transferred graphene, polymer-free methods [175] that transfer graphene without using a polymer supporting layer have also been developed; examples of polymer-free methods include those that use a thin graphite holder (**Fig. 6**b)[176], a waterproof marker frame [177], or electrostatic force [178].

**4.3.5 Dry transfer of CVD graphene**

Dry transfer normally refers to transferring graphene onto the target substrate without using solutions. Similar to how PMMA is used as a polymer supporting layer in wet transfer, PDMS and thermal release tape (**Fig. 6**c) are often used as polymer supporting layers in dry transfer and can be removed by thermal treatment [43,179–182]. For example, the roll-to-roll method for transferring 30-inch graphene has been successfully developed [168]. Specifically, a thermal release tape was attached to graphene by two rollers, and then the Cu substrate was removed by the etchant. Next, the graphene was transferred to the target substrates via two rollers, and the thermal release tape was released by mild heating. The roll-to-roll method is suitable for transferring CVD graphene onto flexible target substrates on a large scale but easily causes mechanical defects when the CVD graphene is transferred to a rigid substrate. An improved roll-to-roll method involving the use of two hot pressure plates was developed [183]; in this method, CVD graphene can be transferred to either flexible or rigid substrates with relatively few defects. On the basis of differences in the adhesion of graphene to various metals, another improved roll-to-roll method involving the use of a thin metal film as the supporting layer of graphene and a thermal release tape was developed [184]; both sides of graphene on the Cu substrate can be transferred, and the Cu substrate can be reused. The dry release transfer of monolayer and bilayer graphene by using poly(propylene) carbonate (PPC) films that have thermoplastic properties according to the adjustable temperature adhesion between the graphene and PPC was demonstrated[185]. Recently, a dry, sacrificial-layer-free and scalable method for transferring graphene and other 2D materials and heterostructures via an adhesive matrix was reported; in this method, the low-adhesion substrate of interest is embedded in a matrix that promotes transfer through strong adhesive interactions with the 2D material[186]. In essence, to realize dry transfer, the adhesion energy between the graphene and substrate should be measured to precisely control delamination.

**4.3.6 Other transfer methods for CVD graphene**

Ultraclean monolayer and bilayer CVD graphene membranes up to 500 μm and 750 μm in size have been realized via the inverted floating method (**Fig. 6**d) followed by thermal annealing under vacuum[187], in which PMMA is used as the supporting layer. A generic methodology for the large-area transfer of graphene and other 2D materials via adhesive wafer bonding was developed[188]; this methodology avoids manual handling and is compatible with large-scale semiconductor manufacturing lines. A simple method of transferring large areas (up to A4-size sheets) of CVD graphene from copper foils onto a target substrate was developed by using a polyvinyl alcohol polymer and hot-roll office laminator[189]. Methods for the direct synthesis, etching and transfer of large-scale and patterned graphene films have been developed[39].

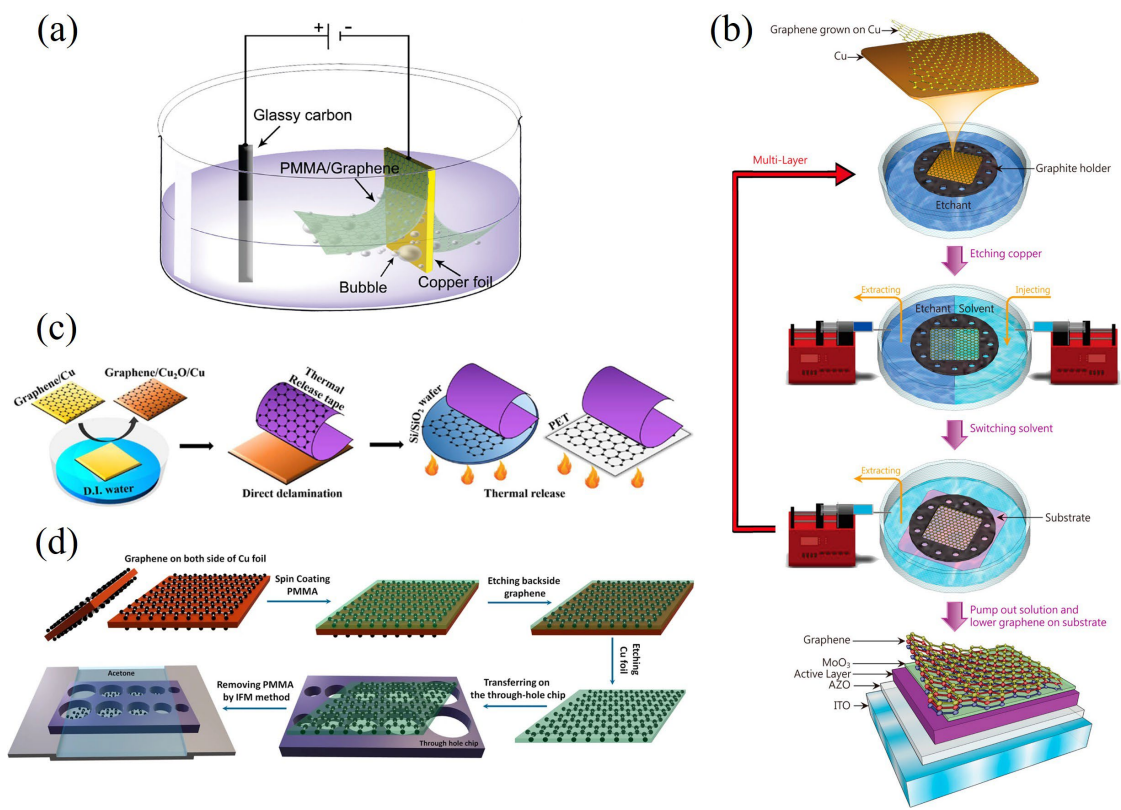

**Fig. 6 Schematics of different methods for graphene transfer. a** Electrochemical exfoliation of CVD-grown graphene[171]. **b** A polymer-free transfer method for CVD-grown graphene[176]. **c** A water-free transfer method for CVD-grown graphene[181]. **d** An improved PMMA-based wet transfer of CVD-grown graphene via the inverted floating method[187]. Figures reproduced with permission from Refs. 171, 176, 181 and 187.

Recently, several new graphene transfer methods have been developed. For example, facile transfer of graphene onto polyethylene without the use of a supporting layer was developed according to van der Waals interactions[190]. A general wafer-scale graphene transfer method based on gradient surface energy modulation was reported; this method results in reliable adhesion and a reliable release of graphene onto target substrate wafers[191]. The large-area transfer of 2D materials, including graphene, was developed by controllable conformal contact[192]; the transferred materials had the advantages of being free of cracks, contamination and wrinkles. More approaches for transferring large-area graphene can be found in the previously reported literature[193–196].

As CVD graphene is transferred from its metallic growth substrate to a target substrate, the technology is compatible with semiconductor fabrication technology at the back end (BEOL). Nevertheless, unintentional contamination by water molecules, lithographic residues, and hydrocarbon contaminants during structuring and contact deposition can significantly influence the electronic system of graphene and modulate its charge carrier density. Notably, these contaminants remarkably limit the sensitivity and applicability of graphene-based sensors.

### 4.3.7. Fabrication of suspended graphene structures

Generally, to fabricate graphene NEMS structures, realizing freely suspended graphene (membranes, beams or ribbons) is typically required (**Fig. 7**a-e)[18,197,198]. Depending on the methods of graphene synthesis, such as CVD on metal foils, mechanical/liquid phase/thermal exfoliation, chemical reduction of graphene oxide and epitaxial graphene on SiC, three main approaches to fabricating suspended graphene include direct mechanical exfoliation of pristine graphene onto prepatterned substrates with trenches or cavities, etching of the substrates underneath the graphene in a final processing step via HF-based wet or dry etching, and transferring CVD-grown graphene directly onto prepatterned substrates with trenches or cavities.

Early suspended graphene sheets were produced by mechanical exfoliation, which restricted scalable fabrication and standardized processes. CVD-grown graphene is a good candidate for fabricating suspended graphene sheets on a large scale via a standardized lithography process. When a wet etching process, such as liquid HF etching, is employed to release the graphene, special care must be taken to avoid the collapse of the suspended graphene by capillary forces and to avoid peeling off the electrodes. The common methods to avoid such collapse are to employ buffered hydrofluoric (BHF) etching followed by critical point drying (CPD) or to employ HF vapor. However, the realization of suspended graphene via HF-based etching also has several drawbacks; for example, HFs, especially liquid BHFs, easily attack Ti, which is often used as a metal electrode material for Ohmic contact with graphene. In addition, the $SiO_2$ underneath the metal electrodes is also easily etched to some extent due to the isotropic etching of BHF, which might affect the stability of the mechanical structures. Thus, finding alternative ways to replace HF etching-based methods to realize suspended graphene structures is meaningful. There are several methods for suspending graphene on organic polymer substrates by using e-beam lithography-based processes to remove parts of the sacrificial polymer layer, thereby avoiding the use of aggressive BHF etching processes (**Fig. 7**a,e)[104,199,200]. In the fabrication process of suspended graphene devices, more attention needs to be paid to cleaning because traditional cleaning methods, such as ultrasonic-assisted dissolution or oxygen ashing, are too aggressive for graphene [201].

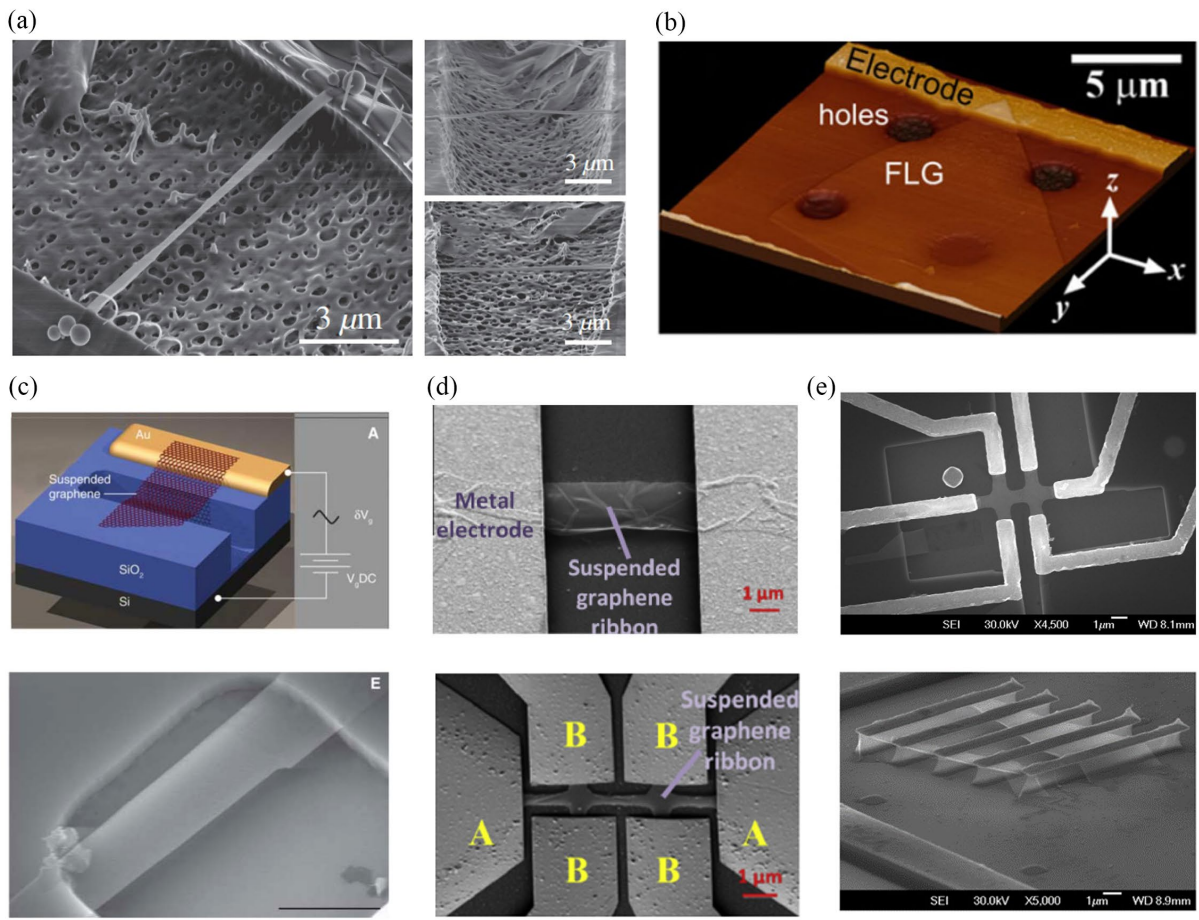

**Fig. 7 Fabrication of suspended graphene structures. a** A few-layer graphene suspended over an etched SU-8 trench for a high-quality factor NEMS resonator due to the thermal shrinkage of the SU-8-induced strain[199]. **b** An AFM topography image of freely suspended few-layer graphene connected to an electrode[197]. **c** Monolayer graphene suspended over an etched SiO$_2$ cavity for the NEMS resonator[14]. **d** SEM images of a graphene ribbon suspended between metal electrodes[201]. **e** SEM images of a high-mobility freely suspended graphene Hall bar that was fabricated on a polydimethylglutarimide-based organic polymer[200]. Figures reproduced with permission from Refs. 14, 197, 199, 200 and 201.

## 5. Patterning and electrical contact of graphene

### 5.1 Patterning of graphene

Typically, graphene must be patterned after it is transferred from the initial substrate to a target substrate for device applications. Although graphene is intrinsically nanoscale, it can be patterned at the wafer scale by a standard lithographic process and $O_2$ plasma etching, which is the most commonly used method for patterning graphene, where a hard or a resist mask is employed[153,202]. For patterning the graphene flakes derived from mechanical exfoliation, standard optical lithography is unavailable, but e-beam lithography or shadow masks are alternative methods. Recently, there have also been reports associated with the laser patterning of graphene after transfer (**Fig. 8**a, b) [203–206] and laser direct writing of graphene patterns on target substrates (**Fig. 8**c)[207,208]. The method for patterning graphene via a laser is flexible and can avoid the possible contamination from photoresist, which the standard lithography process requires. Nanoscale patterning of graphene is also achieved via various methods, such as femtosecond laser ablation [209], local catalytic etching via atomic force microscopy equipped with a Ag-coated probe [210], and focused ion beams, such as helium ion beams [211]. In addition, patterned graphene films can also be directly grown on patterned metal growth substrates (such as Ni layers via CVD) and then transferred to target substrates (such as $SiO_2$ substrates) without additional lithography processes [43].

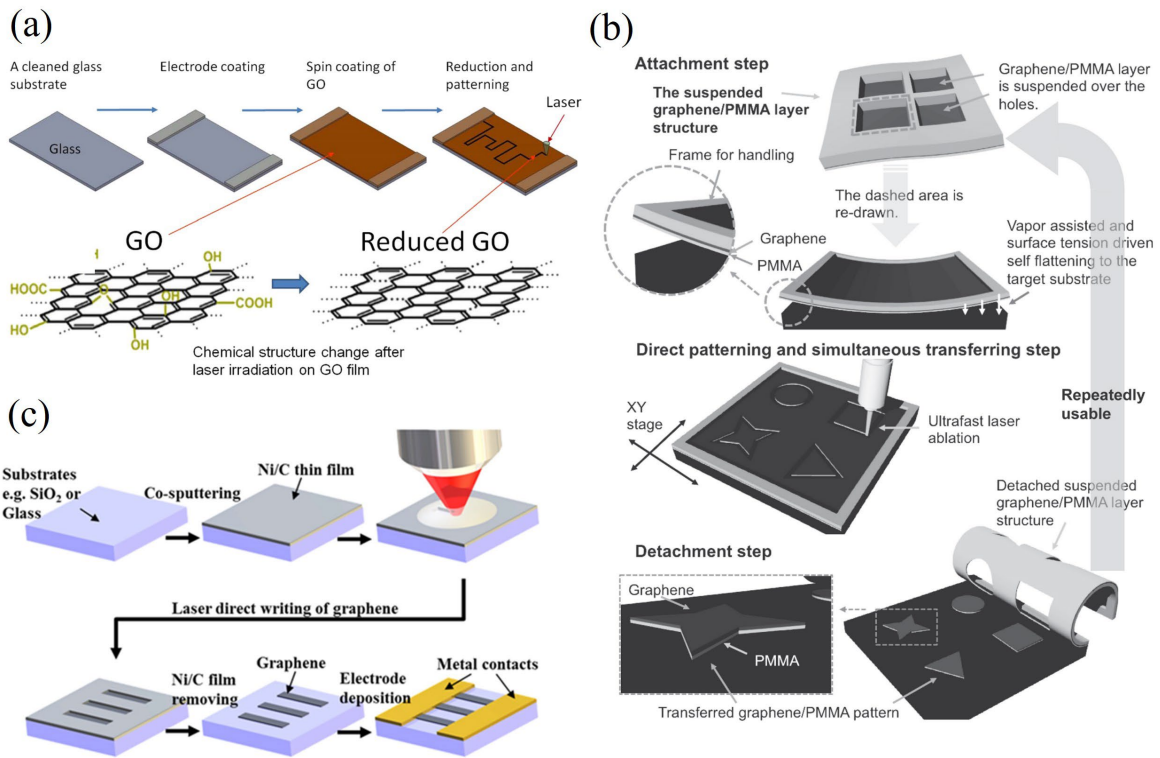

**Fig. 8 Patterning of graphene. a** Laser direct patterning of a reduced-graphene oxide [205]. **b** Laser-induced direct graphene patterning and simultaneous graphene transfer [206]. **c** A laser direct writing process to pattern graphene on insulating substrates under ambient conditions[208]. Figures reproduced with permission from Refs. 205, 206 and 208.

## 5.2 Electrical contact of graphene

Like patterning, graphene must be electrically contacted after the graphene is transferred from the initial substrate to a target substrate for device applications. The metal electrodes for electrically contacting graphene can be fabricated by thermal evaporation or sputtering either before or after graphene transfer. As shown in previous reports [70–72,153,202,212] on graphene-based devices, such as humidity and gas sensors, pressure sensors and accelerometers, before graphene is transferred to a prefabricated $SiO_2$/Si substrate, a 50 nm thick layer of Ti followed by a 270 nm thick layer of Au is evaporated into the pre-etched cavities of the SiO$_2$ layer, thus leaving the patterned Au electrodes to protrude ~20 nm above the SiO$_2$ surface (**Fig. 9**a)[72]. One of the advantages of fabricating contact electrodes before graphene transfer is that the cleanliness of the graphene surfaces is improved and the risk of rupturing the graphene membranes during processing is reduced. Additionally, the graphene–metal contacts are not easily degraded by polymer residues in this way. For a chip package, gold wires are normally bonded from the package to the metal contact pads. The contact resistance between graphene and metal electrodes is crucial for high-performance graphene devices. One review of the literature[213] is specifically associated with studies on the graphene–metal contact characteristics in terms of the metal preparation method, asymmetric conductance, annealing effect and interface impact. There are also two review studies [214,215] related to electrical contact with one- and two-dimensional nanomaterials. As shown in **Fig. 9**b-e[216–219], the contact resistance of graphene devices can be improved by edge-contacted metal–graphene interfaces, the theoretical and experimental study of which has been reported [216–220]. For example, enhanced carrier injection is experimentally achieved in graphene devices by forming cuts in the graphene within the contact regions. These cuts are oriented normal to the channel and facilitate bonding between the contact metal and carbon atoms at the graphene cut edges, thereby reproducibly maximizing "edge-contacted" injection [219].

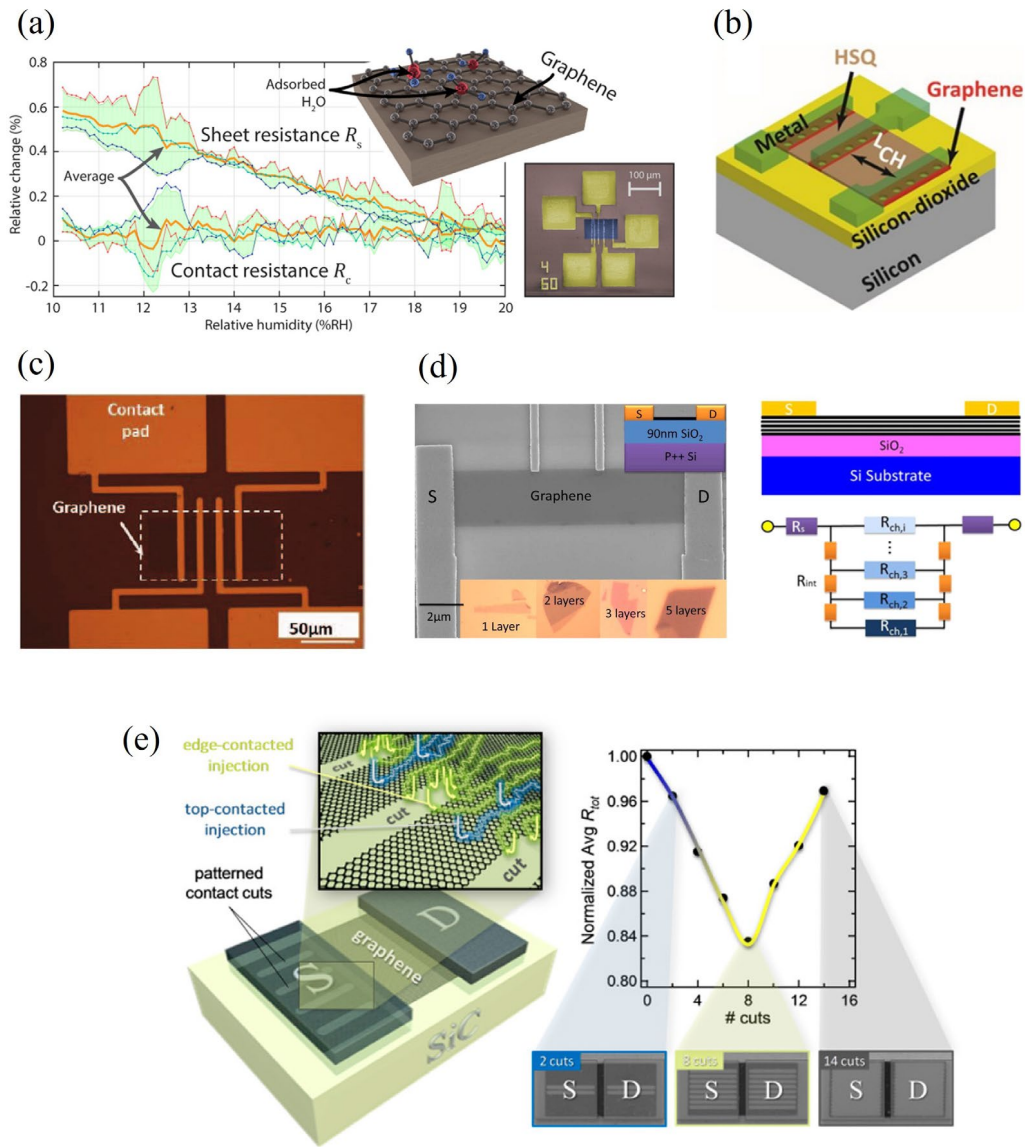

**Fig. 9 Electrical contact of graphene. a** Gold−graphene contact resistance and sheet resistance of graphene versus relative humidity[72]. **b** The graphene under the electrodes was intentionally etched to create holes of different sizes to reduce the contact resistance[216]. **c** Structures used for measuring metal–graphene contact resistances[217]. **d** An SEM image of a four-terminal few-layer graphene device that was used to study the electrical impact of edge contacts and a resistor network model of conventional top-contacted few-layer graphene[218]. **e** A two-terminal graphene device with cuts patterned perpendicular to the channel in the contact region; these cuts can reduce contact resistance[219]. Figures reproduced with permission from Refs. 72, 216-219.

## 6. NEMS devices based on suspended graphene

The earliest study on suspended graphene membranes for NEMS applications in 2007 involved doubly clamped electromechanical resonators that were made of mechanically exfoliated graphene flakes suspended over trenches with thicknesses ranging from monolayer to multiple layers [14]. The fabricated devices represent the thinnest mechanical resonators ever produced and can be actuated by optical or electrical methods in the megahertz range. Following this initial breakthrough, various types of NEMS resonators based on doubly clamped suspended graphene beams and fully clamped suspended graphene membranes have been developed; these resonators include graphene NEMS resonators that can sense mass and temperature [221]. In 2013, the first graphene NEMS pressure sensor with high sensitivity, a small footprint and electrical readout capabilities was realized by suspending a monolayer CVD graphene membrane over the etched cavity [34]. Current applications of suspended graphene in NEMSs have extended from resonators[222,223] and pressure sensors[224] to other types of NEMS devices[23,225,] such as strain sensors [226], nanoelectromechanical switches [110], earphones [227], microphones [112], loudspeakers [111], accelerometers [153,202], and bolometers[228–230]. In this section, we discuss the applications of suspended graphene membranes, beams, and ribbons in various types of nanoelectromechanical devices.

**6.1 Piezoresistive pressure sensors**

As shown in **Fig. 10**a, b, graphene pressure sensors based on the piezoresistive effect have been widely reported [34,87,231,232]. Piezoresistive pressure sensors based on suspended graphene membranes with approximately 20 to 100 times higher sensitivity per unit area (this sensitivity is greater than that of conventional piezoresistive pressure sensors, have been demonstrated [34,87]. In addition, by employing the intrinsic piezoresistivity of graphene to transduce its motion, a suspended H-shaped monolayer graphene resonator clamped at the base by four gold electrodes and SiO$_2$ was fabricated by underetching SiO$_2$ via buffered hydrofluoric acid and critical point drying; this resonator can be used for sensing mass and force with high sensitivity (**Fig. 10**c) [233]. Piezoresistive graphene pressure sensors normally require the impermeability of suspended graphene membranes over cavities. One of the challenges in bringing these sensors into volume production is realizing a hermetically sealed cavity with a reference pressure that remains constant for several years. Although it has been claimed that graphene is impermeable[30], the leakage rates of graphene-sealed cavities are relatively high compared with those of commercial pressure sensor cavities. In addition, obtaining piezoresistive graphene pressure sensors is challenging because of the defects, such as holes and cracks, in suspended graphene membranes.

In addition to pressure sensors based on self-suspended graphene, several pressure sensors use partly suspended graphene with polymers or other membranes [81,234,235]. For example, by transferring few-layer CVD-grown graphene onto a suspended silicon nitride thin membrane that was perforated by a period array of micro through-holes, graphene-based pressure sensors were realized with a sensitivity of $2.8 \times 10^{-5}$ mbar$^{-1}$ and good linearity over a wide pressure range [81]. Another piezoresistive pressure sensor based on multilayer CVD graphene meander patterns located on the maximum strain area of a suspended 100 nm thick, 280 μm wide square silicon nitride membrane was reported [78].

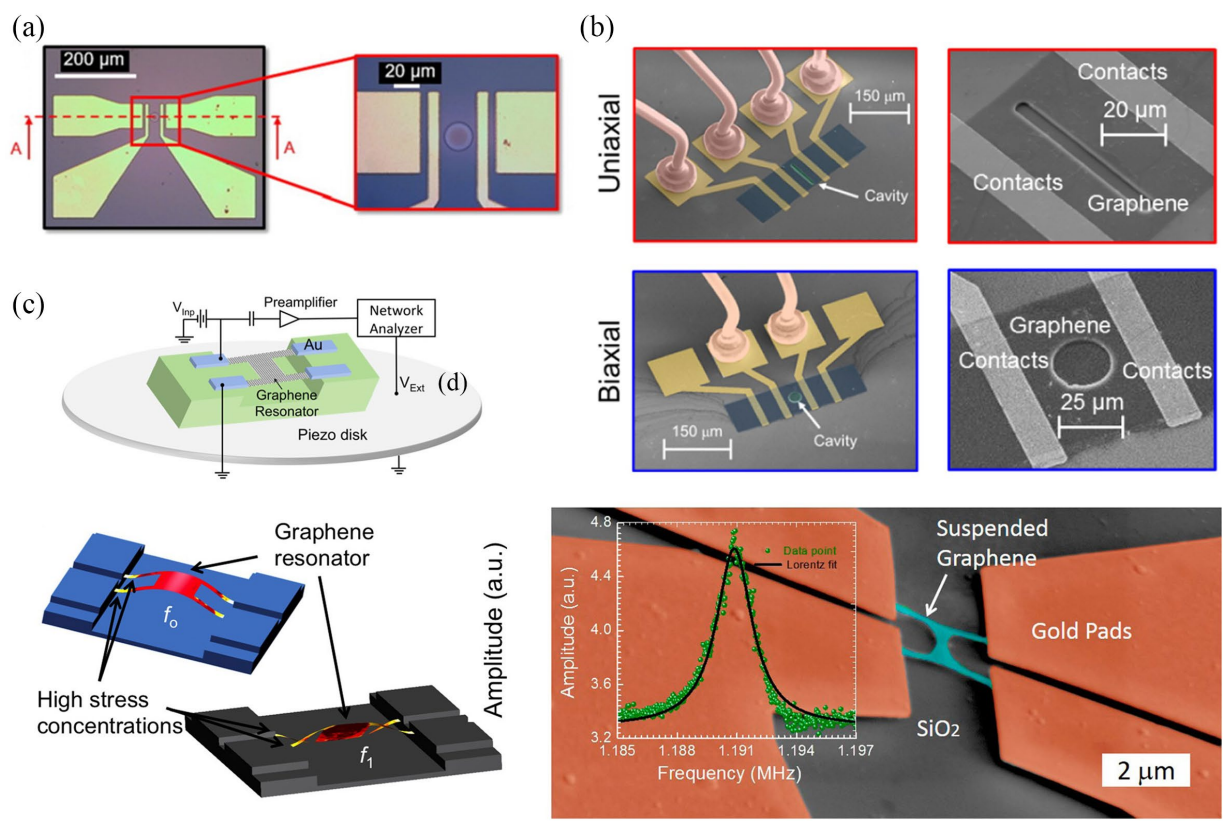

**Fig. 10 Piezoresistive pressure sensors and transducers. a** An optical micrograph of a suspended graphene membrane-based piezoresistive NEMS pressure sensor[231]. **b** SEM images of piezoresistive NEMS pressure sensors based on monolayer graphene suspended over an etched rectangular $SiO_2$ cavity and an etched circular $SiO_2$ cavity[87]. **c** Ultrasensitive room-temperature piezoresistive transduction in a graphene NEMS. This figure part shows a circuit diagram of the measurement setup, a finite element model, and an SEM image of a fabricated suspended H-shaped graphene resonator[233]. Figures reproduced with permission from Refs. 87, 231 and 233.

## 6.2 Capacitive pressure sensors

Graphene-based capacitive pressure sensors have been reported (**Fig. 11**a-d)[94,236,234,237,238]. For example, ultralarge suspended graphene membranes with few-layer graphene over a circular hole with a diameter of up to 1.5 mm on a quartz substrate were realized via a hydrogen bubbling transfer approach with thermal annealing; these graphene membranes were used for capacitive pressure sensors in an impermeable chamber; the results revealed a linear response and a high sensitivity of 15.15 aF Pa$^{-1}$, which is 770% greater than the sensitivity of conventional silicon-based membranes [94]. Another capacitive pressure sensor fabricated on top of an insulating quartz substrate consisted of circular AuPd bottom and AuPd top electrodes that were separated from the bottom electrode by a spin-on-glass dielectric layer, and a few-layer graphene membrane was suspended over the sealed cavity of the AuPd top electrode[236]. The capacitance changes of a graphene drum as low as 50 aF and pressure differences as low as 25 mbar were successfully measured.

In addition to capacitive pressure sensors based on self-suspended graphene, capacitive pressure sensors have been realized by using a suspended ultrathin CVD-grown graphene/PMMA heterostructure membrane over a large array of microcavities, each up to 30 μm in diameter, thus showing reproducible pressure transduction under static and dynamic loading up to pressures of 250 kPa and a pressure sensitivity of 123 aF Pa$^{-1}$ mm$^{-2}$ over a pressure range of 0–100 kPa[234,235].

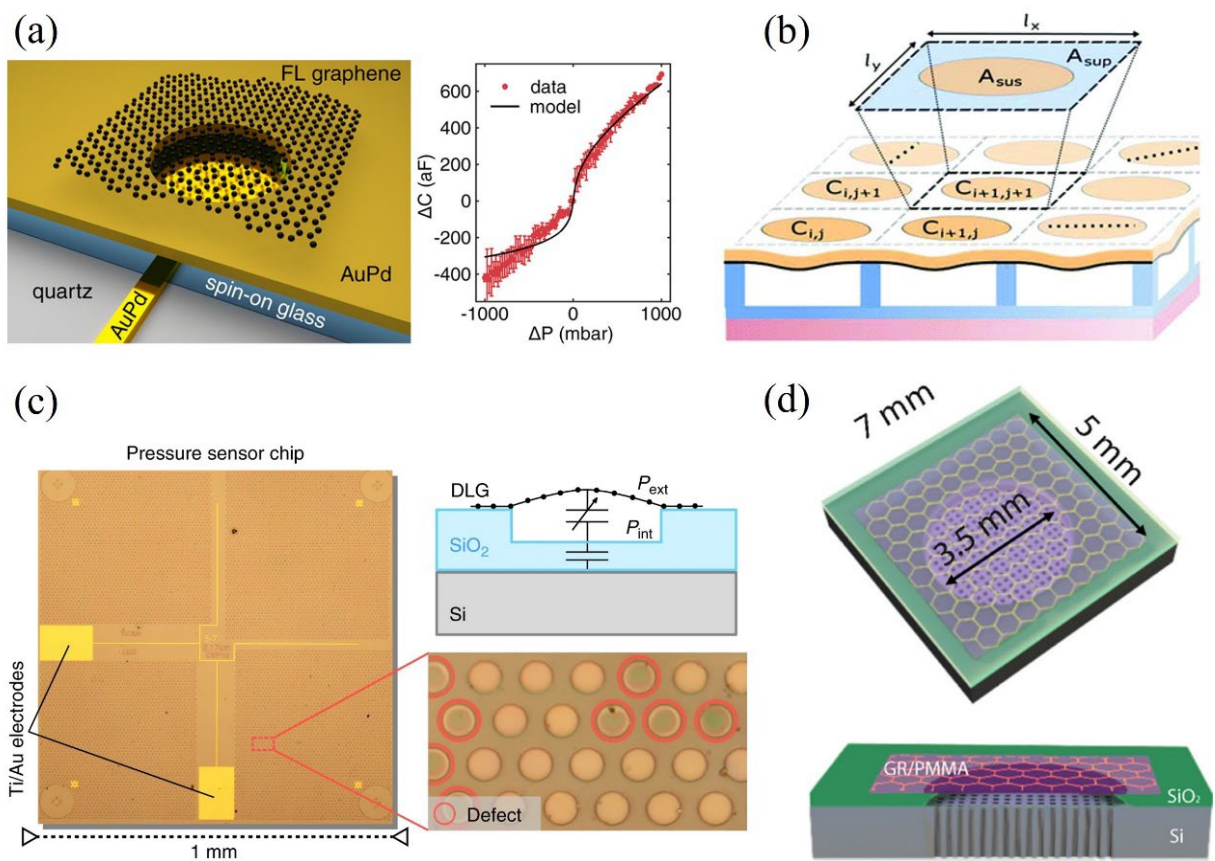

**Fig. 11 Capacitive pressure sensors based on suspended graphene. a** A static capacitive pressure sensor using a single graphene drum[236]. **b** A schematic of a capacitive pressure sensor array using suspended graphene–polymer heterostructure membranes[234]. **c** Sensitive capacitive pressure sensors based on suspended double-layer graphene/PMMA membrane arrays[237]. **d** A schematic of a suspended graphene/PMMA acoustic capacitive sensor created by etching the $SiO_2$ sacrificial layer[238]. Figures reproduced with permission from Refs. 234, 236-238.

## 6.3 Resonant pressure and mass sensors

Graphene pressure sensors based on the resonance effect have been developed [239–242]. For example, squeeze-film pressure sensors are based on the pressure dependence of the membrane frequency; this pressure dependence is caused by the compression of the surrounding gas, thereby changing the stiffness of the resonator, which normally requires strong, flexible and light membranes. As shown in **Fig. 12**a, via a laser interferometry setup for detecting the resonance frequency of the graphene drum, pressure sensors using a few-layer graphene membrane as a squeeze film to cover a gas cavity were demonstrated to have a frequency shift of 4 MHz between 8 and 1000 mbar and a sensitivity of 9000 Hz/mbar; although these results were 45 times greater than that of state-of-the-art MEMS-based squeeze-film pressure sensors, the pressure sensors shown in **Fig. 12**a used a 25-fold smaller membrane area [239]. In addition, **Fig. 12**b shows that resonant pressure sensors using suspended multilayer exfoliated graphene flakes as a sealed drum with low tension were fabricated on insulating sapphire substrates with a local back gate that was used to directly actuate radio frequencies actuation and detect the mechanical resonance modes, thus resulting in high gate tenability (~1 MHz/V) of the resonator modes [241]. In these resonant sensors, the shifts in the resonator mode frequency with the force exerted by the helium gas molecules above the membrane were studied, and the sensing capability of 1 Torr pressure was demonstrated in a cryogenic environment [241].

The equation of the resonance frequency $f = \frac{1}{2\pi}\sqrt{\frac{k}{m}}$ shows that a small mass change[243–245] Δm causes a large frequency shift $\Delta f = -\frac{\Delta m}{2m}f$ because of the high resonance frequency f and low mass m of graphene membranes. Therefore, suspended graphene membranes can be very suitable for mass sensing[246,247]. One of the challenges of mass sensors based on suspended graphene membranes is that the detected shift in resonance frequency is determined by both the mass and position of the mass. [248]The mass of a particle can be determined independently of position by analyzing the frequency shift of multiple well-characterized vibrational modes[249].

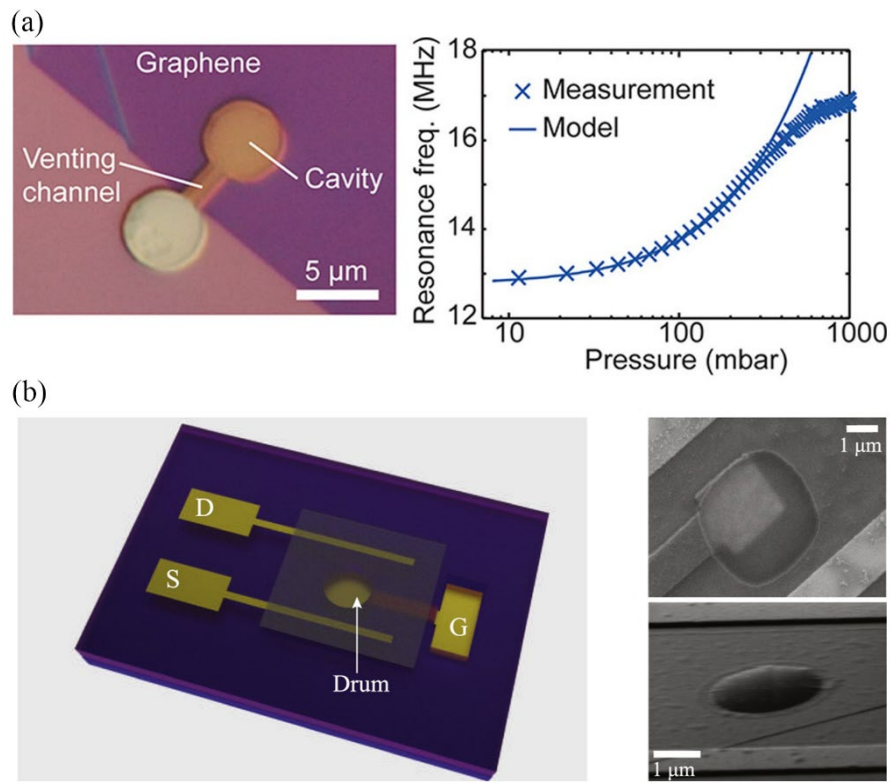

**Fig. 12 Resonant pressure sensors based on suspended graphene. a** A graphene squeeze-film pressure sensor[239]. **b** Graphene drums with low tension for electromechanical resonant pressure sensing[241]. Figures reproduced with permission from Refs. 239 and 241.

### 6.4 Nanoelectromechanical switches

Nanoelectromechanical switches can help reduce power consumption, but reliable nanoelectromechanical switches with low operating voltages and low contact resistances are still challenging and easily fail because of irreversible switching or stiction. Nanoelectromechanical switches employing suspended graphene as movable elements have been developed because of their high Young's modulus, extremely low mass and outstanding carrier mobility and are divided into 2-terminal and 3-terminal NEMS switches [100–104,106,108–

[110,250]. The on and off states of the switches are normally controlled by modulating the electrostatic force applied to graphene. A 2-terminal switch operates by deflecting a suspended graphene membrane with a source–drain voltage and measuring the current once contact. A 3-terminal NEMS switch uses third electrodes to apply an actuation voltage independent of the source-drain voltage and provides merits such as greater operational flexibility, lower power consumption and a higher level of integration.

An early electromechanical switch with two CVD-grown graphene electrodes was realized; this switch could switch several times at a low operation voltage (~ 4.5 V) until the top graphene beam remained in permanent electrical contact with the bottom graphene film [100]. A low pull-in voltage of 1.85 V and a very fast estimated switching time of 40 ns were realized in NEMS switches based on a doubly clamped suspended few-layer graphene beam with a 150 nm air gap[101]. A low pull-in voltage of less than 2 V in graphene NEMS switches based on a simple bottom-up procedure using a polymer sacrificial spacer rather than acid etching was also realized [104]. In addition, an electromechanical thermal switch based on flexible, collapsible graphene membranes with a low operating voltage of approximately 2 V and a thermal switching ratio of approximately 1.3 was reported in 2021[251].

Most of the abovementioned graphene NEMS switches are based on doubly clamped beams and commonly suffer reliability problems because of tears on open edges and/or irreversible stiction of the graphene. To avoid the occurrence of tears in the graphene and to limit stiction, a large array of circularly clamped graphene NEMS switches with a unique design and sub5 V actuation voltage were fabricated via a bottom-up process [108]. They work with either 2-terminal or 3-terminal switching and have line contact during switching to reduce the contact area, thereby holding the potential to address the stiction challenges [108]. Compared with the structures of two-end fixed beams, the structures of cantilevers might have better linear behavior and high sensitivity. Both two-terminal and three-terminal NEMS switches based on suspended graphene cantilever beams were demonstrated with pull-in voltages ranging from 5 V to 10 V [106]. In addition, nanocrystalline graphene directly deposited on insulating substrates can also be used for fabricating high-performance NEMS switches with top electrode contacts on a large scale; these switches exhibit a low pull-in voltage below 3 V, reversible operation, a minimal leakage current of 1 pA, and a high on/off ratio of $10^5$ [110]. A high on/off ratio of $10^5$ was also realized in a graphene NEMS switch with high reproducibility on the basis of locally defined nanomembrane structures in the graphene films on a SiC substrate with atomic steps [102].

## 6.5 Microphones and loudspeakers

In addition to being used for graphene electrostatic switches, graphene is used for electrostatic loudspeakers and microphones[252–256]. Graphene is electrically conductive, has an extremely small mass density, has a high mechanical strength and can construct mechanical resonators, all of these characteristics make graphene an excellent building material for small, efficient, high-quality broadband electrostatic audio speakers[163,254,255,257–264]. On the one hand, the low mass of graphene due to its atomic thickness ensures the high-frequency response of electrostatic audio speakers. On the other hand, the exceptional mechanical strength of graphene allows for relatively large and thin suspended diaphragms, thus reducing the effective spring constant that is required for an effective low-frequency response from electrostatic audio speakers.

A robust speaker built from a multilayered graphene diagram has an excellent frequency response across the entire audio frequency range (20 Hz–20 kHz), with performance matching or surpassing that of commercially available audio earphones [111,112]. Multilayer graphene was used as a membrane material for condenser microphones; with a greater than 10 dB increase in sensitivity, these microphones outperform high-end commercial nickel-based microphones over a significant part of the audio spectrum [265]. A highly sensitive microphone for a hearing aid using a graphene-PMMA laminated diaphragm was created [266]. The graphene-based thermoacoustic loudspeakers [267,268] are especially effective in the ultrasonic frequency region because of the small heat capacity of graphene, in which graphene works as a stationary heater to heat the surrounding air, thereby generating time-dependent pressure vibrations, such as sound waves via thermal expansion (**Fig. 13**a).

### 6.6 Other NEMS devices based on suspended graphene without an attached mass

Suspended graphene can also be used in other aspects of NEMS. For example, a Pirani pressure sensor based on a multilayer suspended graphene strip was realized; the sensor featured a substantially decreased footprint and lower power consumption (**Fig. 13**b)[269]. A noninvasive local optical probe system consisting of a multilayer graphene resonator that is clamped to a gold film on an oxidized silicon surface can be used for quantitatively measuring motion and stress within an NEMS, based on Fizeau interferometry and Raman spectroscopy[107]. A suspended graphene membrane-based gas osmometer with a responsivity of 60 kHz mbar$^{-1}$ was demonstrated on the basis of osmotic pressure-induced deformation of graphene membranes that separate two gases at identical pressures [270]. The osmotic pressure was caused by differences in the rates of gas permeation into a cavity enclosed by few-layer graphene, and the deflection of the membrane was measured on the basis of the tension-induced resonance frequency via a laser interferometric technique [270]. Likewise, as shown in **Fig. 13**c, via the laser interferometric technique, a proof-of-principle for using suspended graphene membranes to pump attolitre quantities of gases at the nanoscale was demonstrated by measuring the deformation of each thin few-layer graphene drum suspended over two circular cavities connected by a narrow trench [271]. Both the cavities and the trench were covered by the graphene membrane, and the local electrodes at the bottom of each cavity allowed each membrane to deform separately on the basis of electrostatic forces and allowed gas to flow inside the trench.

Recently, room temperature detection of the individual physisorption of $CO_2$ molecules with suspended bilayer graphene was reported[272]; a unique device was designed to induce tensile strain in the bilayer graphene to prevent its mechanical deflection onto the substrate by electrostatic force. In 2022, bilayer graphene drums were used to measure the nanomotion of single bacteria in aqueous growth environments[273]. In 2023, NEMS temperature sensors based on suspended graphene were realized on the basis of the piezoresistive properties of graphene [274].

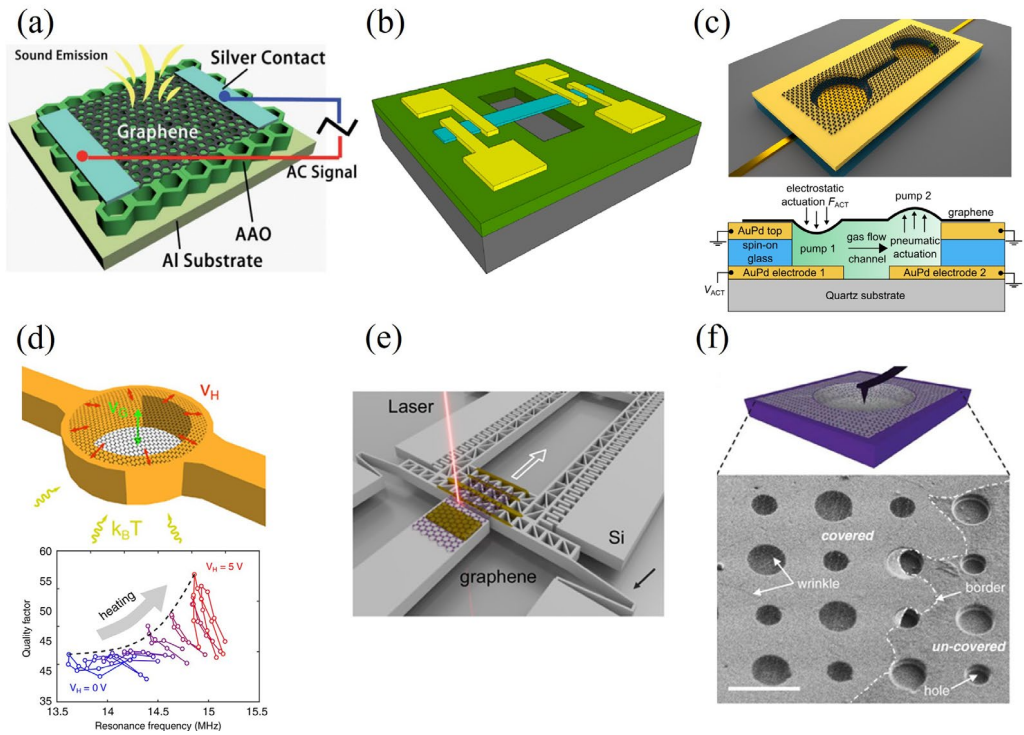

**Fig. 13 NEMS devices based on suspended graphene and characterization of suspended graphene in NEMS structures. a** A schematic view of a monolayer graphene sound-emitting device using anodic aluminum oxide (AAO) templates as substrates[268]. **b** A schematic of a Pirani pressure sensor, in which multilayer graphene suspended over a $SiO_2$ cavity is the sensing element[269]. **c** Graphene gas pumps[271]. **d** On-chip heaters for the tension tuning of graphene nanodrums[275]. **e** A schematic of a comb-drive (CD) actuator that applies uniaxial strain to a graphene flake by moving[276]. **f** AFM nanoindentation experiment of suspended graphene over cavities for measuring Young's modulus[51]. Figures reproduced with permission from Refs. 51, 268, 269, 271, 275 and 276.

## 6.7 Characterization of the suspended graphene in the NEMS structures

NEMS structures can also be used for investigating the properties of graphene[277–284], such as the Young's modulus, tension and strain. An on-chip heater platform for locally tuning the in-plane tension of few-layer graphene nanodrums and investigating the origin of dissipation was reported (**Fig. 14**d)[275]. The thermomechanical resonance frequency and quality factor of graphene can be tuned by increasing the in-plane tension caused by Joule heating of the underlying metallic suspension ring[275]. To avoid any spurious effects due to capacitive coupling to the suspended graphene flake or interaction with the substrate for strain engineering, silicon micromachined comb-drive actuators were fabricated to controllably and reproducibly induce strain in a suspended graphene sheet entirely mechanically (**Fig. 14**e)[276]. In addition, the optimized comb-drive actuator can also be used for strain-dependent transport measurements of suspended graphene [285]. Suspended graphene structures can be used for measuring the Young's modulus of graphene (**Fig. 14**f)[5,51,286–289].

The importance of studying the permeability of suspended graphene in NEMSs lies in their ability to understand the transport mechanisms and interactions of materials at the atomic level and to facilitate the development of new efficient and selective permeable materials. Monolayer graphene with grain boundaries or line defects selectively allows small molecules of gases to pass through. However, large-molecule gases, such as oxygen and nitrogen, can hardly pass through the lattice gaps of graphene[290]. Many advanced methods, such as chemical treatment[291–293], thermal treatment[294], and interlayer modulation[295], have been performed on graphene to modify its permeability. More recently, studies have been conducted to increase the permeability of graphene by introducing nanopores[296]. As a result, the prepared suspended monolayer graphene gas-phase transport molecular valves with discrete angstrom-level pores enable precise control of gas permeability[296].

**6.8 Accelerometers based on suspended graphene with an attached mass**

As discussed above, doubly clamped graphene beams, fully clamped graphene membranes and graphene cantilever beams for applications such as resonators, pressure sensors, switches, and loudspeakers/microphones have been widely studied. However, there are few reports on suspended graphene with attached masses. The limited number of such examples includes micrometre-sized few-layer graphene cantilevers with diamond allotrope carbon weights fabricated by using focused ion beam (FIB) deposition for studying the mechanical properties of graphene (**Fig. 14**a)[297], a kirigami pyramid and cantilevers made of suspended graphene supporting 50 nm thick gold masses that were kept in a liquid to maintain their mechanical integrity (**Fig. 14**b)[298], and a suspended graphene membrane circularly clamped by SU-8 supporting an additional mass made of either SU-8 or gold located at the center of the graphene membrane for shock detection from ultrahigh mechanical impacts (**Fig. 14**c)[299]. All these previous reports involve extremely small masses, and the employed fabrication methods, such as FIB-induced deposition, are slow and typically incompatible with large-scale manufacturing.

Recently, the structures of both suspended graphene membranes and ribbons with large attached $SiO_2$/Si proof masses were realized via a robust, scalable and high-yield MEMS manufacturing approach, and their potential applications in sensitive NEMS piezoresistive accelerometers were demonstrated (**Fig. 14**d, e)[153,202]. The die areas of such graphene-based

NEMS accelerometers are at least two orders of magnitude smaller than those of conventional state-of-the-art silicon accelerometers[153,202]. Ultrasmall and combined spring-mass and piezoresistive transducers based on suspended graphene with an attached mass[202,300–305] are expected to pave the way for a new class of graphene NEMS devices, such as accelerometers with decreased size but increased performance.

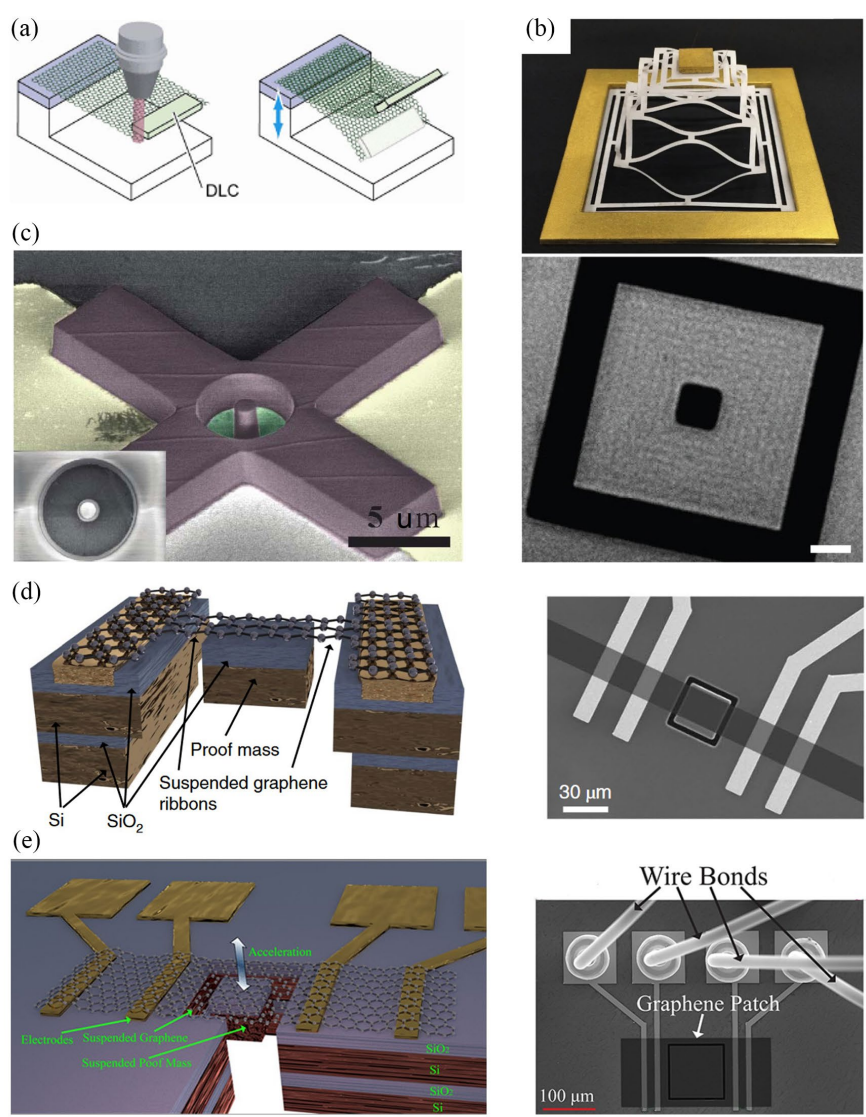

**Fig. 14 Suspended graphene with an attached mass. a** A schematic of a few-layer graphene cantilever with diamond-like carbon weights[297]. **b** A schematic of a kirigami pyramid made of suspended graphene supporting gold masses and an SEM image of the as-fabricated graphene kirigami pyramid[298]. **c** An SEM image of a graphene membrane supporting an SU-8 mass[299]. **d** Doubly clamped graphene ribbons with an attached silicon proof mass as the transducer of the NEMS accelerometers[153]. **e** Fully clamped graphene membranes with attached silicon proof mass as the transducer of NEMS accelerometers[202]. Figures reproduced with permission from Refs. 153, 202, 297-299.

## 7. Devices based on nonsuspended graphene

### 7.1 Flexible transparent conductive electrodes

In addition to the applications of suspended graphene in NEMS, the combination of graphene with conventional membranes also has wide applications in MEMS/NEMS. Owing to their excellent optical transmittance, electrical conductivity, stretchability and mechanical strength, graphene films can be integrated with soft substrates (such as polyethylene terephthalate (PET), PDMS or PET coated with a PDMS layer [43,150,168]) as large-scale stretchable and foldable transparent conductive electrodes for flexible and wearable device applications (including high-speed field effect transistors, touch screens, flat panel displays, light-emitting diodes (LEDs), photovoltaic cells, biomedical sensors and energy-harvesting devices [306–308]). Compared with traditional transparent conductive electrodes, such as indium tin oxide (ITO), although the sheet resistance of graphene films is still not as low as that of ITO, graphene film-based transparent conductive electrodes are strong, flexible, chemically stable and can withstand more strain.

## 7.2 Flexible strain sensors

In combination with its transparency, conformal attachment ability and good stretchability, the sensing features of graphene, such as its piezoresistive properties, make graphene compatible with soft substrates suitable for flexible and wearable sensors, such as strain sensors and pressure sensors. Recently, various types of graphene-based strain sensors that integrate graphene with soft substrates, such as flexible plastic or stretchable rubber substrates (i.e., PDMS), have been widely reported (**Fig. 15**a-c))[66,88,226,309–316]. For example, highly stretchable strain sensors based on graphene–nanocellulose nanopapers were fabricated by embedding three-dimensional macroporous nanopapers composed of crumpled graphene and nanocellulose in a stretchable elastomer (PDMS), which can respond to strains of up to 100% in all directions (**Fig. 15**c)[310]. The strain sensors described above can detect high strain in human-friendly interactive electronics, such as human joint movement with strains of up to 55%[310]. 3D flexible and conductive composites of graphene foams/PDMS were fabricated by growing graphene foams in a nickel foam substrate that was used as a 3D scaffold template, removing the nickel substrate and infiltrating PDMS into the graphene foam, which can be stretched to 95% before the sample is broken [309]. Furthermore, by using nickel foam as the growth template of graphene, immersing graphene-coated nickel foam into the prepolymer of PDMS and chemically etching, the porous graphene network combined with PDMS can be used as both a highly sensitive pressure sensor and strain sensor with a wide measurement range for potential applications in monitoring the walking state, degree of finger bending and wrist blood pressure (**Fig. 15**b) [315]. Nanographene films on PDMS membranes with tunable piezoresistivity for strain sensing with great potential in electronic skin applications can be realized by controlling the nanographene nucleation density and grain size[88].

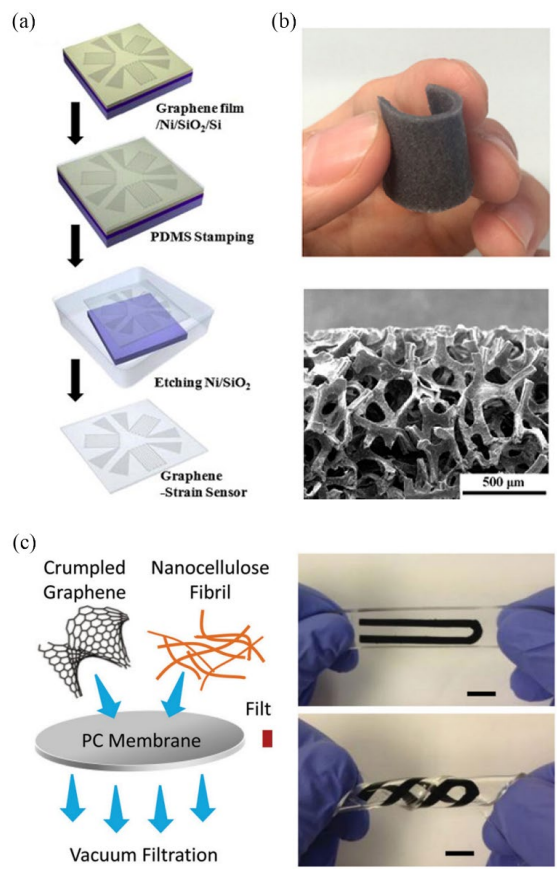

**Fig. 15 Flexible strain sensors. a** The fabrication process of a graphene strain sensor[313]. **b** A photograph of a bent graphene porous network-PDMS composite and SEM image of nickel foam coated with graphene[315]. **c** A stretchable graphene nanopaper[310]. Figures reproduced with permission from Refs. 310, 313 and 315.

### 7.3 Flexible pressure sensors

Different types of flexible and wearable graphene-based pressure sensors that use various types of graphene structures together with other materials or soft substrates as sensing layers have also been reported (**Fig. 16**a, b) [317–320]. For example, pressure sensors based on laser-scribed graphene with a foam-like structure on PET have been reported (**Fig. 16**a)[318]. The sensing principle of a laser-scribed graphene pressure sensor is based on the resistance change between the bottom and top pieces of laser-scribed graphene films that are composed of loosely stacked graphene layers[318]. The sensor sensitivity is as high as 0.96 kPa$^{-1}$ over a wide pressure sensing range (0~50 kPa) because of the large interspacing between graphene layers and the unique V-shaped microstructure of laser-scribed structures[318]. Furthermore, graphene/paper pressure sensors with the ability to measure the wrist pulse, breathing and motion states were realized via the reduction of graphene oxide on the tissue paper in the oven at high temperature, with a pressure range of 20 kPa and ultrahigh sensitivity of 17.2 kPa$^{-1}$ (0–2 kPa) [319]. Graphene/PDMS pressure sensors with a random distribution spinosum microstructure were fabricated by using abrasive paper as a template and the thermal reduction of graphene oxide; the sensitivity of these sensors was as high as 25.1 kPa$^{-1}$ over a wide linear range of 0–2.6 kPa (**Fig. 16**b) [320]. This type of pressure sensor can be used to detect physiological activities, such as human health care monitoring, voice, phonation identification, and motion movement [320].

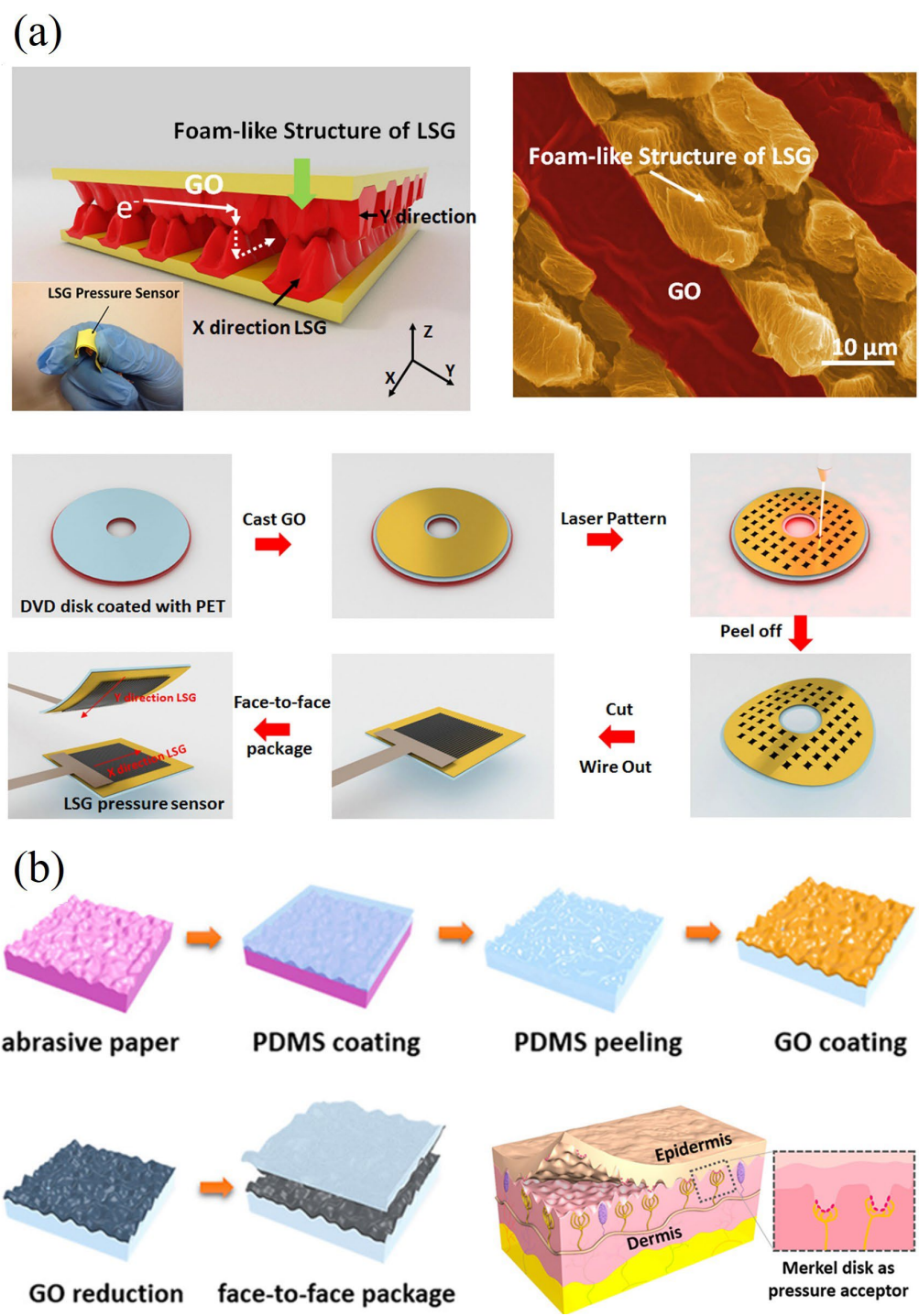

**Fig. 16 Flexible pressure sensors. a** A laser-scribed graphene-based pressure sensor. This figure part shows a schematic and SEM image of the foam-like structure of laser-scribed graphene and the fabrication process[318]. **b** The fabrication process of a bioinspired graphene pressure sensor

with a randomly distributed spinosum microstructure[320]. Figures reproduced with permission from Refs. 318 and 320.

**7.4 Flexible audio-emitting and receiving devices**

Owing to its ultrathin thickness, excellent mechanical properties, superior thermal conductivity and low heat capacity per unit area, graphene can be used for flexible sound generators that can produce sound in audible and ultrasound ranges according to the thermoacoustic effect and has potential applications in speakers, buzzers, earphones, ultrasonic detection and imaging, etc. The basic principle of the thermoacoustic effect can be described as follows: when an alternating current is loaded to a conductor, Joule heating will heat the air near its surface periodically, thereby resulting in the periodic vibration of air and the formation of sound waves. The low heat capacity per unit area of the conductor, which can quickly heat the environment, and the low thermal conductivity of the substrate, which can reduce heat leakage from the substrate, are two key conditions for thermoacoustic sound emission [267]. For example, thermophones made of 20–100 nm thick graphene sheets were demonstrated by patterning graphene sheets on paper substrates[267], thus indicating that graphene sheets have a significant flat frequency response in the wide ultrasonic range of 20 kHz–50 kHz and that thinner graphene sheets can generate higher sound pressure levels. Furthermore, sound-emitting devices based on monolayer graphene were also realized by transferring monolayer graphene onto the substrate of anodic aluminum oxide templates, which therefore reduced heat leakage from the substrate [268].

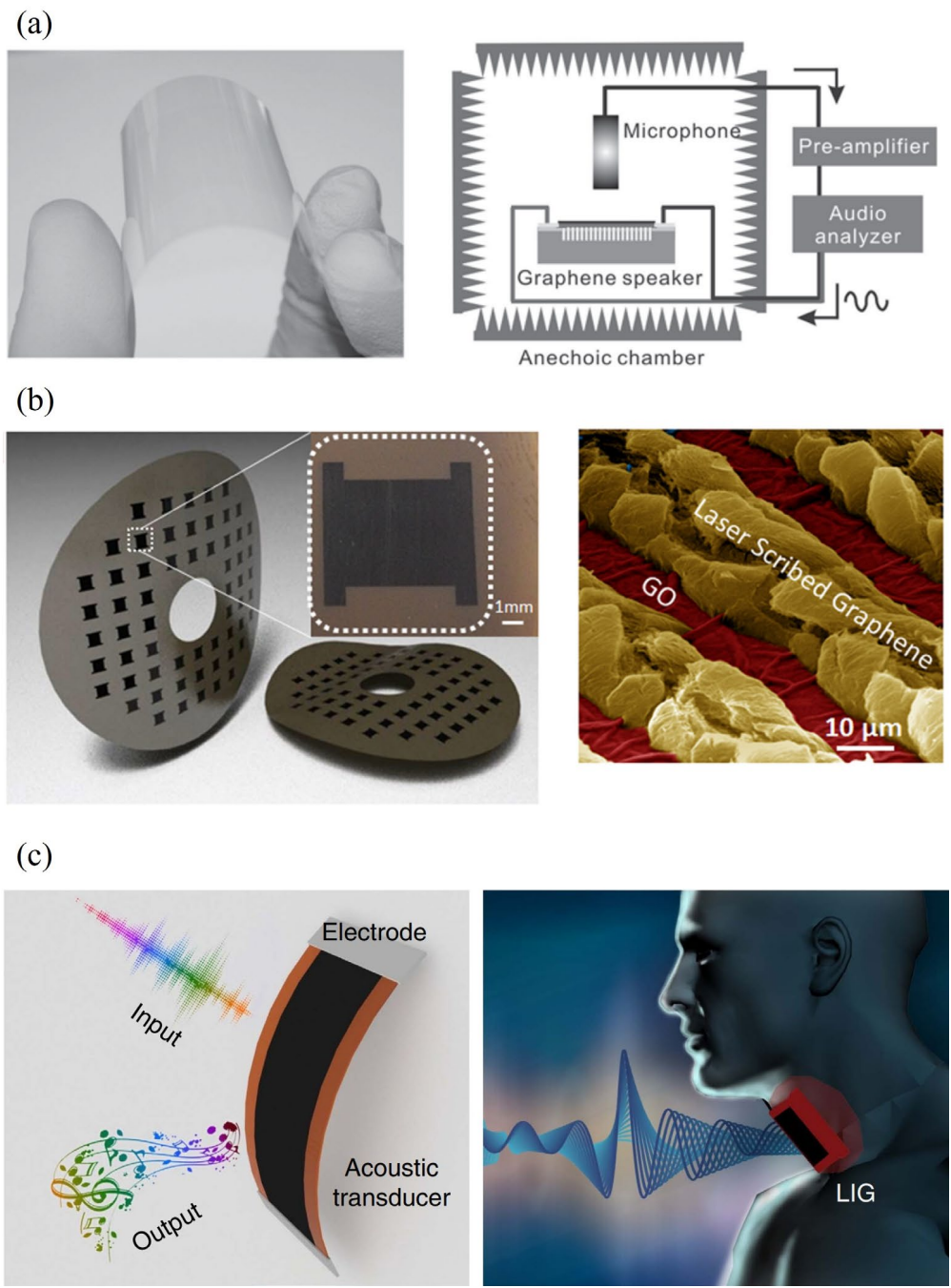

**Fig. 17 Flexible audio-emitting and audio-receiving devices based on nonsuspended graphene. a** Transparent and flexible graphene-based sound sources, including monolayer graphene on PET and a schematic of the acoustic measurement[321]. **b** Wafer-scale flexible graphene earphones and an SEM image of laser-scribed graphene and graphene oxide[227]. **c** An intelligent artificial throat with sound-sensing ability based on laser-induced graphene[322].

Figures reproduced with permission from Refs. 227, 321 and 322.

Combined with the lower heat capacity per unit area of monolayer graphene, monolayer graphene devices can produce higher sound pressures than can graphene sheets. To study the effect of the substrate on sound performance and improve mechanical robustness, double-layer or trilayer graphene was transferred to patterned substrates with different surface porosities, thus indicating that the highest sound pressure was produced from the samples on the substrate with the highest porosity, thereby reducing the heat loss to the supporting substrate as much as possible[321]. In addition to the low heat capacity per unit area of monolayer graphene, the high transmittance and high stretchability of monolayer graphene also allow it to be easily integrated on various transparent supporting substrates, such as PET and PDMS, for transparent and flexible thermoacoustic sound-emitting devices (**Fig. 17**a)[321]. Wafer-scale flexible graphene earphones based on reduced graphene oxide on PET can be realized via laser scribing technology and have a wider frequency response ranging from 100 Hz to 50 kHz, which are suitable for both humans and animals (**Fig. 17**b)[227]. In addition, graphene samples can be used to sense sound according to sound pressure-induced resistance changes of graphene. For example, a flexible intelligent artificial throat made of porous graphene films on a polyimide substrate via laser scribing technology can both generate and detect sound and thus has the potential to assist the disabled (**Fig. 17**c)[322].

**7.5 Flexible actuators**

Actuators based on the combination of graphene with conventional membranes have been studied; these actuators have both potential advantages (including large displacement, rapid response at low voltages, flexibility, patternability and optical transparency) and potential applications (including biomimetic actuators, membranes, and skins). Graphene has a distinctive negative coefficient of thermal expansion (contraction upon heating), which is in contrast with the normal behavior of conventional materials. In addition, graphene has high thermal conductivity. Therefore, a graphene layer can be coupled to different material layers to build microactuators according to the asymmetric thermomechanical response-induced deflection between two different layers upon heating. For example, hybrid bimorph cantilever microactuators based on serpentine graphene sheets pattern/epoxy (SU-8) were realized via batch microfabrication and can be electromechanically driven by applying electrical power, thus resulting in a large displacement of the cantilever, rapid response and low power consumption (**Fig. 18**a)[323]. On the basis of the obvious difference in the coefficient of thermal expansion between graphene paper and graphene oxide paper, an electrothermal hybrid cantilever actuator using graphene and graphene oxide composite paper as building blocks can generate a large displacement with low power consumption by bending the cantilever due to thermal stress during the process of applying an electrical voltage [324].

There are other actuating mechanisms for graphene-based actuators. An actuator composed of multiple stacked graphene electrodes and a dielectric elastomer substrate sandwiched between graphene electrodes showed a displacement of 1050 μm at a frequency of 0.5 Hz with a driving voltage of 2 kV[325]. When a voltage was applied, the area of the dielectric elastomer film with sandwiched graphene electrodes contracted in thickness due to the Maxwell stress because of electrostatic charges and extended along the lateral direction[325]. In addition, bilayer paper samples composed of crisscrossed multiwalled CNTs and graphene oxide platelets were used as the building blocks of microactuators [326], in which graphene oxide could mechanically respond to changes in humidity and/or temperature via changes in the amount of interlamellar water between the graphene oxide platelets. Recently, inspired by origami fabrication, using 2D atomic membranes as a folding material, ultrathin microsized graphene–glass bimorph actuators based on strain control mechanisms of temperature or electrolyte concentration were made by bonding CVD graphene sheets to nanometer-thick layers of glass (**Fig. 18**b)[327].

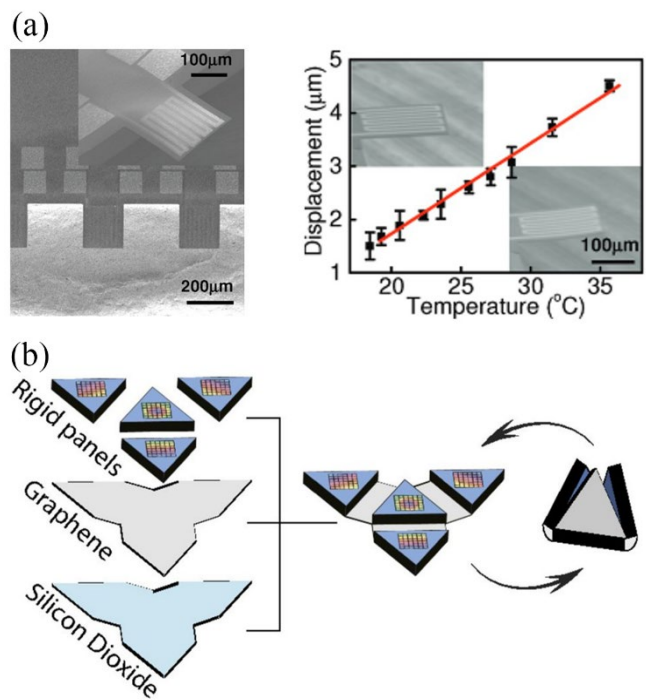

**Fig. 18 Flexible actuators based on nonsuspended graphene. a** A graphene-based bimorph microactuator with a graphene serpentine pattern and an epoxy hybrid microcantilever[323]. **b** The basic structure of a graphene–glass bimorph that is a sheet of graphene bound to a 2 nm thick

layer of glass[327]. Figures reproduced with permission from Refs. 323 and 327.

**7.6 Hall sensors**

Graphene is expected to have the potential to outperform currently available Hall sensor technologies; for example, the current-related sensitivities of graphene-based Hall sensors are much higher than those of state-of-the-art Hall sensors made from silicon and comparable to those of state-of-the-art Hall sensors made from III/V semiconductors[328–330]. The current-related sensitivities of Hall sensors based on large-area unprotected CVD graphene on SiO$_2$ substrates and epitaxial graphene on SiC substrates reach 1200 V/AT [120] and 1020 V/AT[331], respectively. In contrast, the current-related sensitivities of common silicon-based Hall sensors are 6.4 V/AT [332] and 123 V/AT [333].

The graphene Hall devices can present very good linearity over a wide range of magnetic fields, particularly in the voltage range [120]. In addition, the carrier density and mobility of graphene are very weakly affected by temperature; thus, graphene Hall sensors can work linearly over very wide temperature ranges, such as from 1.8 K to 400 K, with high sensitivity and low thermal drift[120], thus substantially widening the application range of graphene Hall sensors. The graphene Hall elements can be manufactured on a large scale at low cost by developing methods for growing and transferring high-quality graphene, can be easily integrated with integrated circuits for signal amplification and processing, and can be fabricated on different substrates, all of which make graphene Hall elements potential substitutes for conventional Hall elements.

To avoid the possible contamination of graphene devices from the surrounding environment and to obtain reliable, durable and high-performance materials for practical applications, the encapsulation of graphene should be considered (**Fig. 19**a-c)[334]. For example, graphene can be encapsulated by the growth of an $Al_2O_3$ layer via atomic layer deposition, thus resulting in improved results with low doping levels [335]. Furthermore, graphene can be encapsulated by 2D insulating h-BN, which has a superior interface, a low amount of dangling bonds and charge traps by van der Waals assembly [36], which are beneficial for high-performance Hall sensor elements. In addition to top encapsulation, h-BN is also very suitable for use as a substrate of graphene in Hall sensor elements for improved performance because of its smoothness. Current-related sensitivities of approximately 1986 V/AT, 2270 V/AT and 5700 V/AT were obtained from batch-fabricated CVD graphene on CVD h-BN substrates without top encapsulation [336], CVD graphene patches encapsulated between exfoliated h-BN [124] and stacks of all-exfoliated h-BN/graphene/h-BN[337], respectively. For more practical applications, graphene-based Hall sensors should be manufactured by using all-CVD-grown 2D material heterostructures in a scalable process to fully encapsulate the active region of graphene (**Fig. 19**c)[338]. In addition to the improved sensitivities, the high linearity [120,123], low noise [123,125,126], wide temperature ranges, thermal stability [120] of the devices, and the transparency and flexibility [124,127,] also make graphene attractive for use in magnetic Hall sensors.

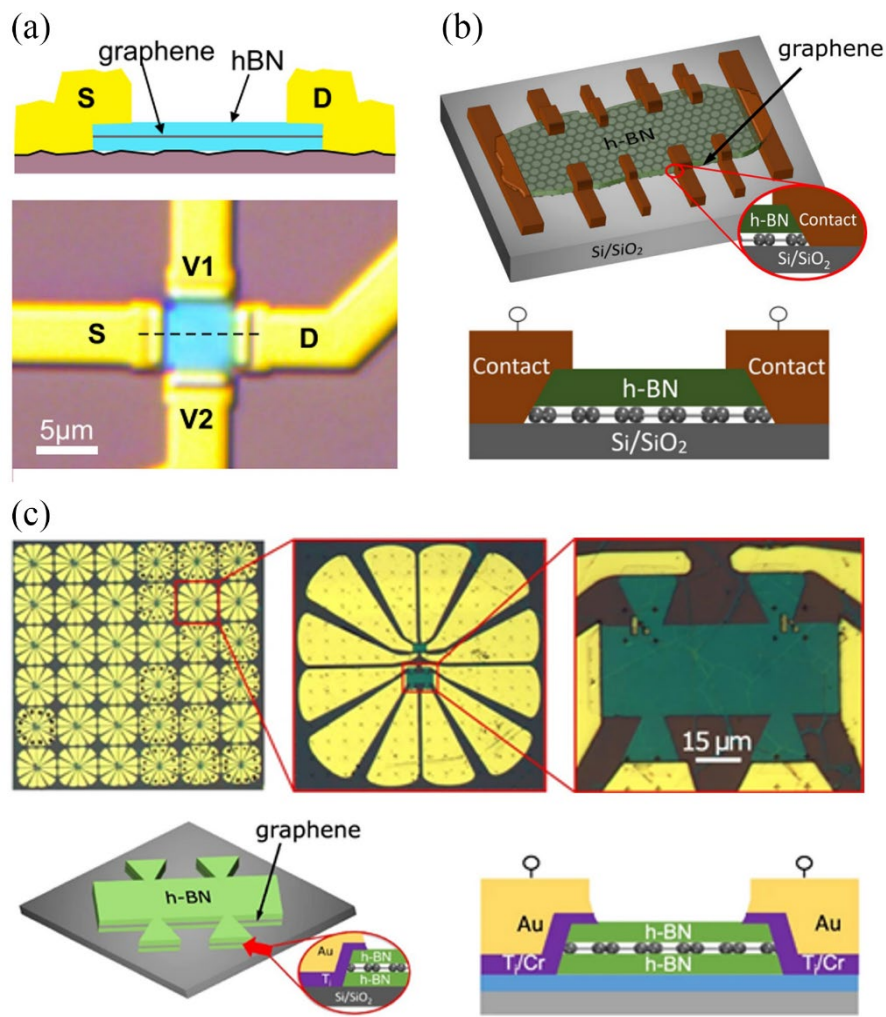

**Fig. 19 Hall sensors. a** A schematic of a cross-section and an optical image of a Hall sensor based on graphene encapsulated in h-BN[337]. **b** Hall sensors fabricated on large-area CVD graphene protected by h-BN with 1D edge contacts[334]. **c** A batch-fabricated chip and an individual Hall element with an h-BN/graphene/h-BN heterostructure[338]. Figures reproduced with permission from Refs. 334, 337 and 338.

## 8. Conclusions and outlook

The quality of graphene materials and their commercial availability will mature during the coming years[339–342], thus opening up new exciting applications of graphene in MEMS and NEMS and sensor applications, including flexible and stretchable devices for wearable sensors[43,308]. Graphene is thinner than any currently available CMOS or MEMS/NEMS material and thus could substantially reduce device size while increasing sensitivity. In addition to graphene, many 2D materials (e.g., graphane, fluorographene, h-BN, $MoS_2$ and other transition metal dichalcogenides (TMDs)) with various exceptional properties, such as from electrically insulating to conductive and from strong to soft, are emerging. For example, some 2D materials have interesting electrical and mechanical properties, such as piezoelectric behavior and a strong piezoresistive effect in $MoS_2$-based TMDs [343–345]. Although the research on emerging 2D materials is not mature and a number of 2D materials are chemically unstable under atmospheric conditions, some of these 2D materials have already been used for sensing applications [346,347]. These emerging 2D materials also have great potential for use in MEMS and NEMS and sensor applications in the future, with the expectation of a substantial reduction in the sizes of the NEMS sensors and improved sensitivity. Furthermore, heterostructures of 2D materials[348,349] and sandwich structures made of two, three, or more economical layers of different 2D materials can provide more opportunities to optimize the properties and structures of materials with atomic precision, possibly yielding more possible applications, including various future MEMS/NEMS and sensor applications.

In this review, we provide an overview of the demonstrated NEMS sensors and proof-of-concept devices based on graphene. These devices are generally smaller than their conventional MEMS counterparts. Furthermore, they showed improved performance and sometimes even completely novel functionalities. However, there are still some challenges for graphene-based

NEMS sensors. We believe that one of the greatest challenges that graphene-based NEMS sensors face is the relatively low-yield manufacturing capabilities of suspended graphene structures. Especially for suspended monolayer graphene structures, the yield is generally quite low. We have provided an overview of potential manufacturing approaches. In addition to different manufacturing approaches that might result in a low yield of suspended graphene, the low yield can essentially be attributed to the grain boundaries and related defects in CVD-grown graphene, which easily result in the fracture of suspended graphene. To improve the yield of the suspended graphene, the CVD method for growing graphene should be continuously improved so that sufficiently large grains of graphene with as few grain boundaries as possible can be realized. Other solutions for improving the yield of suspended graphene include the continuous improvement of graphene transfer methods, the use of BHF etching followed by CPD or the use of HF vapor to release graphene for suspension, and the use of two to ten atomic layers of graphene or stacks of graphene with polymers or other membranes.

Another of the greatest challenges that graphene-based NEMS sensors face is their relatively poor stability and reliability. The electrical signal from practical graphene-based NEMS sensors, especially monolayer graphene-based devices, is extremely sensitive to various environmental parameters, such as humidity, gas, temperature, and electromagnetic fields, thus resulting in relatively poor stability and reliability. To address this challenge, graphene should be carefully shielded from environmental influences. For example, graphene can be encapsulated by other membranes, such as h-BN, $Si_3N_4$, $Al_2O_3$, or PMMA-based polymers.

In addition, developing a feasible approach for manufacturing multiple types of suspended sensors in a single fabrication flow is highly important and is a typical challenge. Other challenges include high-performance electronic readout circuits, packaging, and reliability measurements for graphene-based NEMS devices.

**Acknowledgments**

This work was supported by the National Natural Science Foundation of China (62171037 and 62088101), the 173 Technical Field Fund (2023-JCJQ-JJ-0971), the Beijing Natural Science Foundation (4232076), the National Key Research and Development Program of China (2022YFB3204600), the Beijing Institute of Technology Science and Technology Innovation Plan (2022CX11019), the National Science Fund for Excellent Young Scholars (Overseas), the Beijing Institute of Technology Teli Young Fellow Program (2021TLQT012), and the FLAG-ERA project 2DNEMS funded by the Swedish Research Council (VR) (2019-03412).**Conflict of interest**

The authors declare no competing interests.